\documentclass[fleqn,usenatbib]{mnras}

\usepackage{newtxtext,newtxmath}

\usepackage[T1]{fontenc}
\DeclareRobustCommand{\VAN}[3]{#2}
\let\VANthebibliography\thebibliography
\def\thebibliography{\DeclareRobustCommand{\VAN}[3]{##3}\VANthebibliography}

\usepackage{graphicx}	
\usepackage{amsmath}	
\usepackage{lastpage}
\usepackage{color}
\usepackage{subfigure} 
\usepackage{rotating}

\title[Elemental abundance variations in M31]{Cloud-scale elemental abundance variations and the CO-to-dust-mass conversion factor in M31}

\author[C. Bosomworth et al.]{
Chloe Bosomworth$^{1,2}$\thanks{E-mail: cb20acq@herts.ac.uk},
Jan Forbrich$^{1,2}$,
Charles J Lada$^{2}$,
Nelson Caldwell$^{2}$,
Chiaki Kobayashi$^{1}$,\newauthor
Sébastien Viaene$^{3}$ 
\\
$^{1}$Centre for Astrophysics Research, University of Hertfordshire, College Lane, Hatfield, AL10 9AB, UK\\
$^{2}$Center for Astrophysics | Harvard \& Smithsonian, 60 Garden Street, MS 72, Cambridge, MA 02138, USA\\
$^{3}$Sterrenkundig Observatorium, Universiteit Gent, Krijgslaan 281, B-9000, Gent, Belgium \\
}

\date{Accepted XXX. Received YYY; in original form ZZZ}

\pubyear{2024}

\begin{document}
\label{firstpage}
\pagerange{\pageref{firstpage}--\pageref{LastPage}}
\maketitle

\begin{abstract}

From a spectroscopic survey of candidate H~{\sc ii} regions in the Andromeda galaxy (M31) with MMT/Hectospec, we have identified 294 H~{\sc ii} regions using emission line ratios and calculated elemental abundances from strong-line diagnostics (values ranging from sub-solar to super-solar) producing both Oxygen and Nitrogen radial abundance gradients. The Oxygen gradient is relatively flat, while the Nitrogen gradient is significantly steeper, indicating a higher N/O ratio in M31’s inner regions, consistent with recent simulations of galaxy chemical evolution. No strong evidence was found of systematic galaxy-scale trends beyond the radial gradient. After subtracting the radial gradient from abundance values, we find an apparently stochastic and statistically significant scatter of standard deviation 0.06 dex, which exceeds measurement uncertainties. One explanation includes a possible collision with M32 200 - 800 Myrs ago. Using the two-point correlation function of the Oxygen abundance, we find that, similar to other spiral galaxies, M31 is well-mixed on sub-kpc scales but less so on larger (kpc) scales, which could be a result of an exponential decrease in mixing speed with spatial scale, and the aforementioned recent merger. Finally, the MMT spectroscopy is complemented by a dust continuum and CO survey of individual Giant Molecular Clouds, conducted with the Submillimeter Array. By combining the MMT and SMA observations, we obtain a unique direct test of the Oxygen abundance dependence of the $\alpha^{\prime}(^{12}{\rm CO})$ factor which is crucial to convert CO emission to dust mass. Our results suggest that within our sample there is no trend of the $\alpha^{\prime}(^{12}{\rm CO})$ with Oxygen abundance.

\end{abstract}




\section{Introduction}

H~{\sc ii} regions are clouds of gas that have been photoionized by nearby massive stars, the OB stars, often located near the edges of Giant Molecular Clouds (GMCs).
OB stars have a short lifetime of $\lesssim 10$ Myr, therefore the elemental abundances of H~{\sc ii} regions can be used to trace recent star formation and thus the latest stages of a galaxy's chemical evolution. Elemental abundance ratios such as Oxygen (O) and Nitrogen (N) to Hydrogen (H) i.e. O/H and N/H (e.g., \citealp{Pilyugin_2010, SanchezMenguiano_2016}) as well as N/O (e.g., \citealp{Belfiore_2017}) when traced across an entire galaxy, can be used to investigate its star formation history. Both star formation and the chemical history of galaxies are critical to our understanding of galaxy formation and evolution \citep{Henry_1999}. 

Stars form from the ISM and nucleosynthesis gives rise to an increasing abundance of heavier elements  (e.g., O and N) which are returned to the ISM upon death of the star. In disk galaxies, the resulting trend is dominantly a linear, radial decline in H~{\sc ii} region metallicity because of the higher star-formation rates in the inner galaxy (\citealp{Tinsley_1980}). Accretion of external low metallicity gas dilutes these products of nucleosynthesis and thus also affects the composition of the ISM.

Early on, measurements of gas-phase metallicity gradients have helped build analytic models of chemical evolution for galaxies \citep{Matteucci_1989}. These gradients have also been compared with chemodynamical simulations of galaxies (e.g., \citealp{Vincenzo_2020}). This topic has recently been summarised in the review papers \citet{Kewley_2019} \& \citet{Maiolino_2019}.
Once this radial gradient is removed from the data and only residual metallicity, $\Delta$(O/H), differences remain, the presence of higher-order differences within the ISM can be explored (e.g., \citealp{Ho_2017}, \citealp{Ho_2018}, \citealp{Kreckel_2019}, \citealp{Williams_2022}).

More recently, the focus has shifted from galactic-scale trends to local variation driven by mixing of the ISM. For example, \cite{Kreckel_2020} identify a high level of chemical homogeneity over large (kpc) spatial scales, demonstrating that efficient mixing within the ISM on these scales is a common feature in their sample of spiral galaxies. On smaller scales, local enrichment of the ISM can occur due to nearby star formation (e.g. \citealp{Ho_2017, Groves_2023}). N-body/hydrodynamical simulations of disc galaxies by \citet{Khoperskov_2023} involving multiple spiral arms, a radial metallicity gradient and local enhancement in the presence of radial migration of gas, predict azimuthal scatter in gas-phase metallicity at a given galactocentric radius up to $\sim$0.05 dex associated with the spiral structure. 

O and N abundances are used as proxies to trace gas-phase chemical abundances of H~{\sc ii} regions throughout galaxies (e.g., \citealp{Pilyugin_2010}, \citealp{Pilyugin_2016}). O, produced by massive stars, has a high relative abundance and therefore produces strong optical emission lines, making it the easiest proxy to measure.
In stellar nucleosynthesis, N is produced from the CNO cycle. In the primary process, the seed C is produced in the same production site e.g., massive (4--8~$M_\odot$) AGB stars and rotating massive stars (see footnote 4 in \citealp{Kobayashi_2023}). In Galactic chemical evolution models (\citealp{Kobayashi_2020} and references therein), N enrichment from AGB stars appears at higher metallicities of the ISM due to their longer timescales than massive stars (the N enrichment is not secondary but just delayed), after the ISM is enriched by massive stars. In the secondary process, N is produced from the CNO already present in the progenitor star, and thus N yield depends on the metallicity of the star \citep{Clayton_1983, Arnett_1996}. The secondary N increases also for metal-rich SNe, which also appears at high metallicities of the ISM. Because of the combination of these, the N/O ratio is roughly proportional to the initial metallicity of the star-forming gas (see \citealp{Vincenzo_2018}). Note that, confusingly, the observed N increase at high metallicity is often called “secondary N”, but is caused by both primary process in AGB stars and the secondary process in massive stars.

\defcitealias{Sanders_2012}{S12}
The Andromeda Galaxy (M31) is our most nearby large spiral galaxy ($\rm \sim 780 pc$: \citealp{Stanek_1998}). It has a prominent, star-forming ring at a galactocentric radius of approximately 10 kpc, seen in infrared (e.g., \citealp{Habing_1984, Gordon_2006}) and $\rm H\alpha$ (e.g., \citealp{Arp_1964, Devereux_1994}) images. The star-forming ring, known as the "Ring of Fire" (RoF), has been shown to split near the position of M32 \citep{Gordon_2006}, displaying a second, fainter ring at $\rm \sim 14~kpc$ galactocentric radius \citep{Haas_1998}. In this paper, we report sub-cloud--scale ($\rm \sim 5~pc$) metallicity measurements toward individual H~{\sc ii} regions in M31. We then performed further analysis with a focus on metallicity differences as a function of spatial separation between H~{\sc ii} regions. Using MMT/Hectospec \citep{Fabricant_2005}, we have conducted a multi-object spectroscopic survey for hundreds of candidate H~{\sc ii} regions in M31. We have used line ratio measurements to both classify sources as H~{\sc ii} regions (distinguishing from planetary nebulae) and calculate their metallicities with strong line diagnostics. A range of metallicities and complex trends have been found in the past for some of the same sources by \citet{Sanders_2012} (hereafter \citetalias{Sanders_2012}). 

The most relevant work to this study, given the identical instrumentation used, is the optical spectroscopic survey of M31 conducted by \citetalias{Sanders_2012}. Also using HECTOSPEC data, they identified planetary nebulae (PNe) and H~{\sc ii} regions in M31 and calculated O and N abundances for these sources from emission line ratios. For PNe, the temperature-sensitive auroral line $\rm [OIII]\lambda4363$ is used to calculate direct O abundances. In H~{\sc ii} regions, this line is too weak to be detected, and so various strong-line diagnostics are used. These are indirect methods of measuring chemical abundances derived from the relations between ratios of strong emission lines and directly measured metallicities from observations (e.g., \citealp{Zaritsky_1994, Pilyugin_2010, Marino_2013}) or from photoionization models (e.g., \citealp{Kewley_2002}) or even a combined calibration approach (e.g., \citealp{PettiniPagel_2004}). \citetalias{Sanders_2012} utilise five different O abundance diagnostics, metallicities were calculated for sample sizes of 48 - 192 H~{\sc ii} regions depending on frequencies of line detections. Radial abundance gradients calculated by \citetalias{Sanders_2012} are in agreement with previous works using the same methods but also depend on the strong-line diagnostic used. For all diagnostics used, \citetalias{Sanders_2012} find significant intrinsic scatter. Using their preferred strong-line method for H~{\sc ii} regions (O3N2 from \citealp{Nagao_2006}), they found that $33\%$ of neighboring (within 0.5 kpc) pairs of H~{\sc ii} regions vary by $> 0.3$ dex.

Planetary nebulae (PNe) observations can also be used to trace galaxy formation and chemical evolution, and this has been done previously for M31 (e.g., \citetalias{Sanders_2012}, \citealp{Bhattacharya_2022}, \citealp{Arnaboldi_22}, \citealp{Kobayashi_2023}). Compared to H~{\sc ii} regions which reflect recent star formation, PNe abundances reflect the older ISM as they are ionized by red-giant stars at the end of their lifetimes. Temperature-sensitive auroral lines such as [OII]$\lambda4363$, [OII]$\lambda7325$ and [SIII]$\lambda6312$ are typically stronger in PNe than H~{\sc ii} regions and thus detected more often. From the ratio of strong-lines in the spectra to the corresponding auroral line, electron temperature ($\rm T_{e}$) can be measured. This allows for their metallicities to be calculated by "direct" methods (e.g., \citealp{Bresolin_2010, Bhattacharya_2022}).
By calculating direct oxygen and argon abundances for $> 200$ PNe, \cite{Bhattacharya_2022} compared the abundance gradients of PNe in the thin and thick disks of M31 and provided evidence that these are chemically distinct. For the thin disk, a significant negative O abundance gradient was found of similar magnitude to those found for M31 H~{\sc ii} regions. In comparison, the thicker disk was found to have a slightly positive radial gradient. One explanation offered was that the thin disk formed following a wet-merger event (a major merger of mass ratio 1:5 occuring $\sim$ 2.5 - 4.5 Gyr ago; \citealp{Bhattacharya_2019}); a collision between two gas-rich galaxies, where metal-poor gas was brought in by the satellite galaxy during the merger, causing a burst of star formation. Simultaneously, the thicker disk was "radially homogenized" due to this wet-merger event \citep{Bhattacharya_2022}.  

The star-formation rate (SFR) of a galaxy depends on the mass of molecular gas ($\rm M_{mol}$) available in the ISM, and thus the mass of GMCs is crucial to investigating galaxy evolution. From the mass-metallicity relation (MZR; \citealp{Lequeux_1979}) we know that gas-phase oxygen abundance increases with a galaxy's total stellar mass $\rm M_{*}$. \citet{Mannucci_2010} studied the relationship between (O/H), $\rm M_{*}$ and SFR, which defines the FMR (Fundamental Metallicity Relation). Scaling relations have also been found to display secondary dependencies at z = 0, the most relevant for this study is the gas-FMR \citep{Bothwell_2016} where at a fixed $\rm M_{*}$, O/H is inversely related to $\rm M_{mol}$. By studying $\rm M_{mol}$ and metallicities of the same GMCs we can begin to investigate these scaling relations.

The primary constituent of GMCs, $\rm H_2$, is not observable in emission and therefore $\rm M_{mol}$ is often measured indirectly using the second most abundant emitter: CO. To convert this to $\rm M_{mol}$, the conversion factor between CO luminosity and $\rm M_{mol}$ ($\alpha_{\rm CO}$) is required. $\alpha_{\rm CO}$ is predicted to depend on metallicity; UV radiation destroys molecules depending on the amount of extinction present, thus predicting that the CO to $\rm H_{2}$ ratio decreases at lower metallicities where UV can penetrate deeper into the clouds (e.g., \citealp{Pelupessy_2009}, \citealp{Shetty_2011}). However, in order to calculate $\alpha_{\rm CO}$ we must assume a value for the gas-to-dust ratio. To avoid this, \cite{Forbrich_2020} derived the conversion factor for CO luminosity to dust mass ($\rm M_{dust}$), $\alpha^{\prime}(^{12}{\rm CO})$, which does not depend on this assumption and can be measured more directly. \cite{Viaene_2021} report $\alpha^{\prime}({\rm CO})$ for $^{12}\rm CO$ and $^{13}\rm CO$ towards 20 GMCs in M31, some of which we obtained metallicity values for associated H~{\sc ii} regions and therefore we directly investigate the dependence of $\alpha^{\prime}({\rm CO})$ on metallicity.  

This paper is structured as follows: in Section \ref{sec: obs} we describe our observations and how these were reduced. In Section \ref{sec: method} we discuss our methodology including emission line flux measurements, extinction correction, source classification and metallicity/abundance measurements. Our results, including radial metallicity gradients and two-point correlation functions of metallicity are presented in Section \ref{sec:results}. We present our summary \& conclusions in Section \ref{sec:sum}. 

\section{Observations and Data Reduction} \label{sec: obs}
\subsection{Source Selection} \label{sec: DC}

Our targets were primarily selected from a catalog of candidate H~{\sc ii} regions, \cite{Azimlu_2011}, many of which are associated with one of the 326 known Giant Molecular Associations (GMAs) in M31 from \cite{Kirk_2015} (from which the previously mentioned SMA targets were selected). Some additional sources from a VLA survey (Toomey et al. in prep) were also observed. GMAs for which we have $\alpha_{\rm CO}$ measurements from SMA data \citep{Viaene_2021, Forbrich_2020} were of the highest priority for observation so that the effect of metallicity on GMC properties can be investigated. Selecting our targets in this way increased the probability that we would observe H~{\sc ii} regions with measurable emission lines.

\subsection{Observations}

Observations were performed using the HECTOSPEC spectrometer \citep{Fabricant_2005} at the MMT telescope located in Arizona, USA, using the 270 gpm grating covering the wavelength range $\sim$3700 - 9150 $\Angstrom$ with a 5 $\Angstrom$ resolution. Within this wavelength we have access to multiple strong emission lines; H$\alpha$, H$\beta$, [OII]$\lambda3727$, [OIII]$\lambda\lambda(4959,5007)$, [NII]$\lambda\lambda(6548,6584)$ and [SII]$\lambda\lambda(6717,6731)$. Observations were taken during two separate observing runs; the first in October 2020 and the second in November and December of 2021. HECTOSPEC has 300 fibres with an on-sky fibre radius of 1.5" (corresponding to $\sim$ 5 pc at the distance of M31), and the entire field is $\sim$ 1 deg in diameter. This was ideal for our purposes as we aimed to obtain a large sample size of H~{\sc ii} regions. For each pointing, these fibres are allocated as follows: on-sky fibres for sky-subtraction (a total of 35 across all runs) (see section \ref{sec:DR}), 12 - 24 pointing fibres (located at the outer edge of the spectrograph), with the rest allocated to science targets. 

\begin{figure}
    \centering
    \includegraphics[width = \columnwidth, trim = {0cm, 1cm, 0cm, 0cm}]{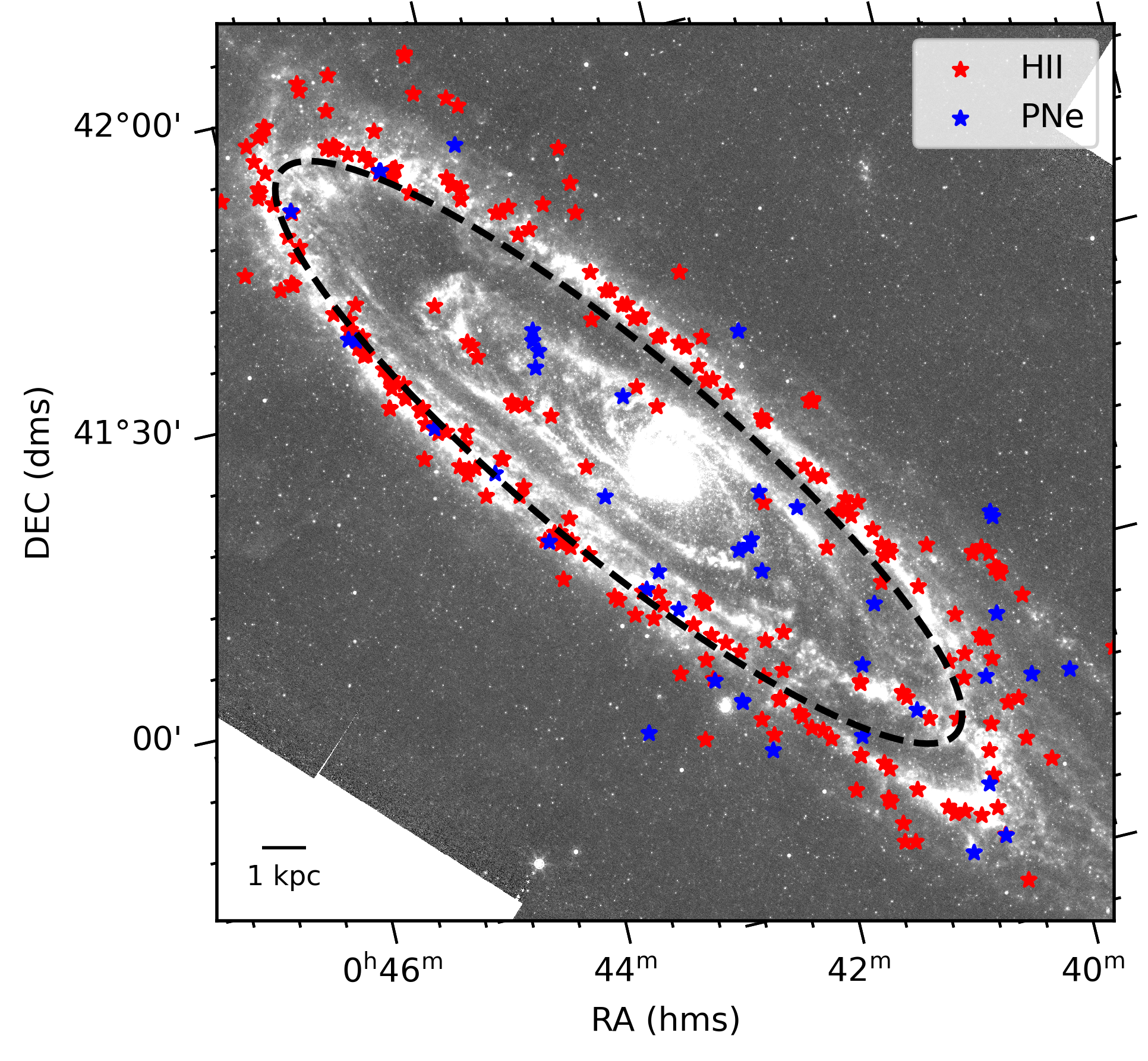}
    \caption{H~{\sc ii} regions (red) and PNe (blue) are shown on the Spitzer MIPS Infrared 24 microns image \citep{Gordon_2006}. We display the main RoF as defined by \citet{Gordon_2006} as a black, dashed line. The 1 kpc scalebar represents this distance on-sky at the distance of M31 (780 kpc: \citealp{Stanek_1998}). These are the targets from which we obtained high-quality Hectospec spectra and were able to classify them using their emission line ratios.}
    \label{fig:DSS}
\end{figure}

We have high-quality spectra for 294 H~{\sc ii} regions in M31, and $\sim$ 300 other sources including 44 PNe. The locations of the H~{\sc ii} regions are shown in Figure \ref{fig:DSS} and were distinguished from PNe as described in section \ref{sec:CLASS}. Our target selection process enabled us to observe the entire disk at the galactocentric radii range $\sim4 - 21$ kpc. Unsurprisingly, we see that the majority of H~{\sc ii} regions in M31 lie within the inner "Ring of Fire" (RoF) at  $\sim$ 10 kpc, or the outer RoF at $\sim$ 14 kpc (\citealp{Haas_1998, Gordon_2006}), where the majority of star formation occurs in M31 \citep{Lewis_2015}. In Figure \ref{fig:DSS}, we show the inner RoF as defined by \citet{Gordon_2006}; a circle of $\sim$ 10 kpc radius, offset from the galaxy centre by ($\rm 5.5^{\prime}, 3.0^{\prime}$).

The majority of our targets were observed multiple times, thus allowing us to confirm measurements from multiple spectra. Also, many \citet{Kirk_2015} GMAs have multiple associated H~{\sc ii} regions in our source sample. As most H~{\sc ii} regions are much larger than 5 pc (e.g., \citealp{Azimlu_2011} sources have diameters up to $\sim$190 pc) we primarily observe portions of H~{\sc ii} regions. Following \cite{Sanders_2012}, we assume that any inhomogeneities within individual H~{\sc ii} regions are small and random and the region observed is representative of the entire H~{\sc ii} region within given uncertainties. 

\subsection{Data Reduction} 
\label{sec:DR}

The observations were reduced using the IDL based Hectospec reduction pipeline `Hsred 2.0' (https://www.mmto.org/hsred-reduction-pipeline/), the steps performed are as follows: The frames were first debiased and flat fielded. Separate exposures were compared to identify and remove cosmic rays through interpolation. Individual spectra were then extracted, combined and wavelength calibrated. A number of fibers in each pointing were assigned to blank areas of sky within the focal plane, these are combined and used to perform sky-subtraction on the target spectra. This sky-subtraction method has already been compared to the use of different sky spectra (from areas of the sky ranging from local to distant) in \citetalias{Sanders_2012} and only small differences were found in emission line ratios (mean difference of $\sim 0.01$ dex) if different sky spectra were used for data reduction.

We applied a flux correction model to all spectra. There was a lack of known flux-calibration sources in our field of view, therefore, we use calibration information obtained from a 2019 observation of the flux standard Feige 34. This calibration is a per bin correction from 2019 with an identical wavelength axis to our spectra. A typical target airmass of 1.3 was assumed. The flux calibration model is identical to that used by \citetalias{Sanders_2012}. For matching sources, we compared our emission line fluxes after flux calibration, but without extinction correction, with their Table 2. We found excellent agreement between the emission line fluxes.

\section{Methodology and Data Analysis} \label{sec: method}

In this section we describe our methodology. Emission line fluxes were measured with the intention of using the emission line ratios to classify sources, perform extinction correction, and calculate metallicities.

\subsection{Emission Line Fluxes}
\label{sec: ELF}

For emission line flux measurements, we utilise the Python package Specutils \citep{Specutils}, applying a method similar to that of \citetalias{Sanders_2012}. Due to uncertainty in the flux calibration, we do not compute absolute values for line fluxes. Following Sanders et al. (2012) and others, we report line fluxes after renormalizing the spectra to $\rm H\beta$ = 100. We extracted line profiles of width 60 \AA~(centered on the midpoint of the line) and subtract a linear continuum (fit to the two 20 \AA~ ranges on both sides of the line profile). We then fit Gaussians to the continuum-subtracted line profiles for the following lines: H$\alpha$, [OII]$\lambda3727$, [OIII]$\lambda\lambda(4959,5007)$, [NII]$\lambda\lambda(6548,6584)$ and [SII]$\lambda\lambda(6717,6731)$, and measure line fluxes. See Appendix \ref{sec: LineProfs} for representative examples. We find line fluxes compatible within uncertainties with those of \citetalias{Sanders_2012} for any matching sources. In cases where lines are within 60\AA~of one another (H$\alpha$ \& [NII]$\lambda\lambda(6548,6584)$ and [SII]$\lambda\lambda(6717,6731)$), we used a multiple Gaussian fit with one peak centered on each emission line in the range. Continuum subtraction is done still assuming a linear continuum fit, the corresponding extracted spectrum contains the required lines with an extension of 20 \AA~ on both sides. 

For each individual target, lines for the same transition were Doppler-shifted due to the blueshift of M31 as a whole and the rotation of the galaxy. We identified an initial velocity guess for each spectrum using the CO velocity field from \citet{Nieten_2006} in order to identify the midpoint of each emission line. In the rare case that no CO velocity value is available within 10 pixels of the observation coordinates, the median CO velocity of the entire field, $-267$ km/s, was used as an initial guess. This method provided accurate enough wavelength values so that the exact line-midpoints count be determined via a Gaussian fit. 

The S/N of the line flux measurements was calculated based on the ratio of the flux value to the rms of the continuum-subtracted noise of 20~\AA~ width on either side of the emission line. We introduced an S/N $>$ 5 detection limit for lines so as to ensure we are detecting only emission lines that are clearly above any noise. Extinction-corrected emission line fluxes were used for metallicity calculation after the initial ratio of $\rm H\alpha/H\beta$ was calculated (see section \ref{sec:EC}). The extinction corrected line flux ratios for our final sample of H~{\sc ii} regions for which we calculate corresponding abundances (see Section \ref{sec:met}) are reported in Table \ref{tab:tab_fluxes}. We show an example line profile with Gaussian fits for each strong-line measured for one of our spectra in Appendix Section \ref{sec: LineProfs}. No correction has been made for underlying stellar absorption as we estimate that this will have a very minimal effect on Balmer line fluxes. A test was already conducted on MMT spectra of H~{\sc ii} regions in M31 by \citetalias{Sanders_2012}. Models from Starburst99 \citep{Leitherer_1999} were subtracted from underlying continua for both a low and high metallicity H~{\sc ii} region. They found that the ratio of $\rm H\gamma/H\beta$ changes by 1\% at the most from uncorrected values. Additionally, they compared the H$\alpha$ equivalent widths with extinction ($\rm A_{v}$) and found no significant correlation. This indicates that Balmer absorption does not significantly affect measured line ratios for our H~{\sc ii} region sample.

While the overall error budget for the emission line flux can be difficult to ascertain, our repeat observations of the same targets provided us with the dispersion of multiple independent measurements for the majority of sources. In the cases in which we have only one observation for a source, the mean percentage uncertainty of this emission line (from sources that do have multiple corresponding spectra) is reported. We introduce a lower bound of 0.1 to uncertainties to reflect unquantifiable systematic uncertainties. 
We compare these to flux uncertainties derived via error propagation from pipeline-produced sigma spectra, and from those based on line S/N, and found in both cases that these errors were significantly smaller than those from averaging individual runs. 

\begin{table*}
\centering
	\caption{Emission line flux ratios (relative to $\rm H\beta = 100$) for M31 H~{\sc ii} regions. Nominal uncertainties are estimated from the standard deviation between values obtained from repeat observations of the same source, where available (see text), with a lower bound of 0.1.}
	\label{tab:tab_fluxes}
	\begin{tabular*}{\linewidth}{@{\extracolsep{\fill}} cccccccccc} 
		\hline
		ID&R.A., Dec.&[OII]&[OIII]&[OIII]&[NII]&$\rm H\alpha$&[NII]&[SII]&[SII]\\ & (J2000) & $\lambda~3727$ & $\lambda~4959$ & $\lambda~5007$ & $\lambda~6548$ & $\lambda~6563$ & $\lambda~6584$ & $\lambda~6717$ & $\lambda~6731$ \\
		\hline
		1 & 0:39:13.01, 40:41:44.88 & 321.6 $\pm$ 25.8 & 26.2 $\pm$ 2.1 & 67.4 $\pm$ 5.2 & 18.5 $\pm$ 1.3 & 291.3 $\pm$ 0.6 & 67.7 $\pm$ 3.1 & 23.7 $\pm$ 1.2 & 18.7 $\pm$ 1.1 \\ 2 & 0:39:14.70, 40:48:33.12 & 459.6 $\pm$ 109.0 & 25.8 $\pm$ 3.5 & 75.0 $\pm$ 34.9 & 20.7 $\pm$ 3.5 & 289.8 $\pm$ 1.3 & 64.5 $\pm$ 4.7 & 31.6 $\pm$ 1.5 & 20.3 $\pm$ 1.9 \\ 3 & 0:39:16.50, 40:41:04.92 & 370.1 $\pm$ 7.2 & 54.3 $\pm$ 1.3 & 159.6 $\pm$ 3.2 & 19.1 $\pm$ 1.4 & 290.3 $\pm$ 0.4 & 56.3 $\pm$ 2.9 & 21.8 $\pm$ 1.1 & 15.9 $\pm$ 0.8 \\ 4 & 0:39:55.39, 40:55:49.18 & 258.0 $\pm$ 40.4 & 13.8 $\pm$ 1.0 & 42.6 $\pm$ 1.5 & 33.0 $\pm$ 1.0 & 293.0 $\pm$ 0.2 & 100.0 $\pm$ 2.6 & 27.4 $\pm$ 0.9 & 19.1 $\pm$ 1.1 \\ 5 & 0:40:00.60, 40:39:12.60 & 357.8 $\pm$ 19.3 & 10.7 $\pm$ 2.6 & 32.0 $\pm$ 3.9 & 40.5 $\pm$ 1.5 & 297.5 $\pm$ 1.0 & 124.6 $\pm$ 4.1 & 26.4 $\pm$ 3.9 & 18.1 $\pm$ 1.9 \\ 6 & 0:40:03.89, 40:58:27.12 & 359.9 $\pm$ 48.2 & 40.9 $\pm$ 3.2 & 126.7 $\pm$ 6.6 & 40.5 $\pm$ 1.3 & 284.9 $\pm$ 0.1 & 119.2 $\pm$ 4.3 & 61.6 $\pm$ 0.9 & 43.3 $\pm$ 3.6 \\ 7 & 0:40:04.01, 40:58:53.39 & 238.9 $\pm$ 1.6 & 41.6 $\pm$ 0.1 & 125.0 $\pm$ 1.9 & 18.9 $\pm$ 0.6 & 284.7 $\pm$ 0.1 & 54.5 $\pm$ 2.9 & 18.8 $\pm$ 0.1 & 13.2 $\pm$ 0.3 \\ 8 & 0:40:04.30, 40:58:45.84 & 407.3 $\pm$ 86.9 & 62.6 $\pm$ 0.9 & 183.6 $\pm$ 11.8 & 38.8 $\pm$ 2.9 & 284.8 $\pm$ 0.1 & 108.9 $\pm$ 3.4 & 49.2 $\pm$ 1.7 & 34.7 $\pm$ 0.8 \\ 9 & 0:40:05.30, 40:59:07.44 & 380.9 $\pm$ 89.8 & 55.7 $\pm$ 5.0 & 164.9 $\pm$ 11.3 & 38.2 $\pm$ 1.9 & 284.8 $\pm$ 0.1 & 100.6 $\pm$ 1.4 & 52.7 $\pm$ 1.2 & 36.4 $\pm$ 1.6 \\ 10 & 0:40:06.36, 40:59:04.05 & 184.9 $\pm$ 25.8 & 71.8 $\pm$ 2.1 & 197.0 $\pm$ 5.2 & 20.9 $\pm$ 1.3 & 284.9 $\pm$ 0.6 & 56.0 $\pm$ 3.1 & 27.4 $\pm$ 1.2 & 18.8 $\pm$ 1.1 \\
        \hline
	\end{tabular*}
 \begin{minipage}{\linewidth}~\\
     The full table of 294 sources is available online, we show a portion here for 10 sources.
 \end{minipage}
\end{table*}

\subsection{Extinction Correction}
\label{sec:EC}
Interstellar extinction varies significantly depending on the line of sight and is dependent on wavelength. Dust attenuation affects key observable properties of star formation in galaxies, including emission line fluxes. To account for any discrepancies between emission line fluxes of separate H~{\sc ii} regions due to dust attenuation, we applied the optical/near-infrared (NIR) extinction curve from \cite{Cardelli_1989} with $R_{v} = 3.1$. Since H~{\sc ii} regions are primarily ionized by photons of energy 13.6 eV which are usually not reabsorbed, case B recombination is assumed along with electron density and temperature typical of H~{\sc ii} regions ($\rm 10^{4}cm^{-2}$ and $\rm 10^{3}K$ respectively). An intrinsic Balmer line ratio of H$\alpha$/H$\beta = 2.86$ is expected \citep{Osterbrock_2006}. We calculated a value for visual extinction ($A_{V}$) for each source following the method of \cite{Momcheva_2013}. Our extinction values were found to be compatible with those of \citetalias{Sanders_2012} for the 20 sources we have in common. Our $A_{V}$ values ranged from 0 - 4.7 with a median value of 1.1. Individual $A_{V}$ for our sample are reported in Table \ref{tab:tab_met}.

\begin{figure}
    \centering
    \includegraphics[width=\columnwidth, trim = {0cm, 1cm, 0cm, 0cm}]{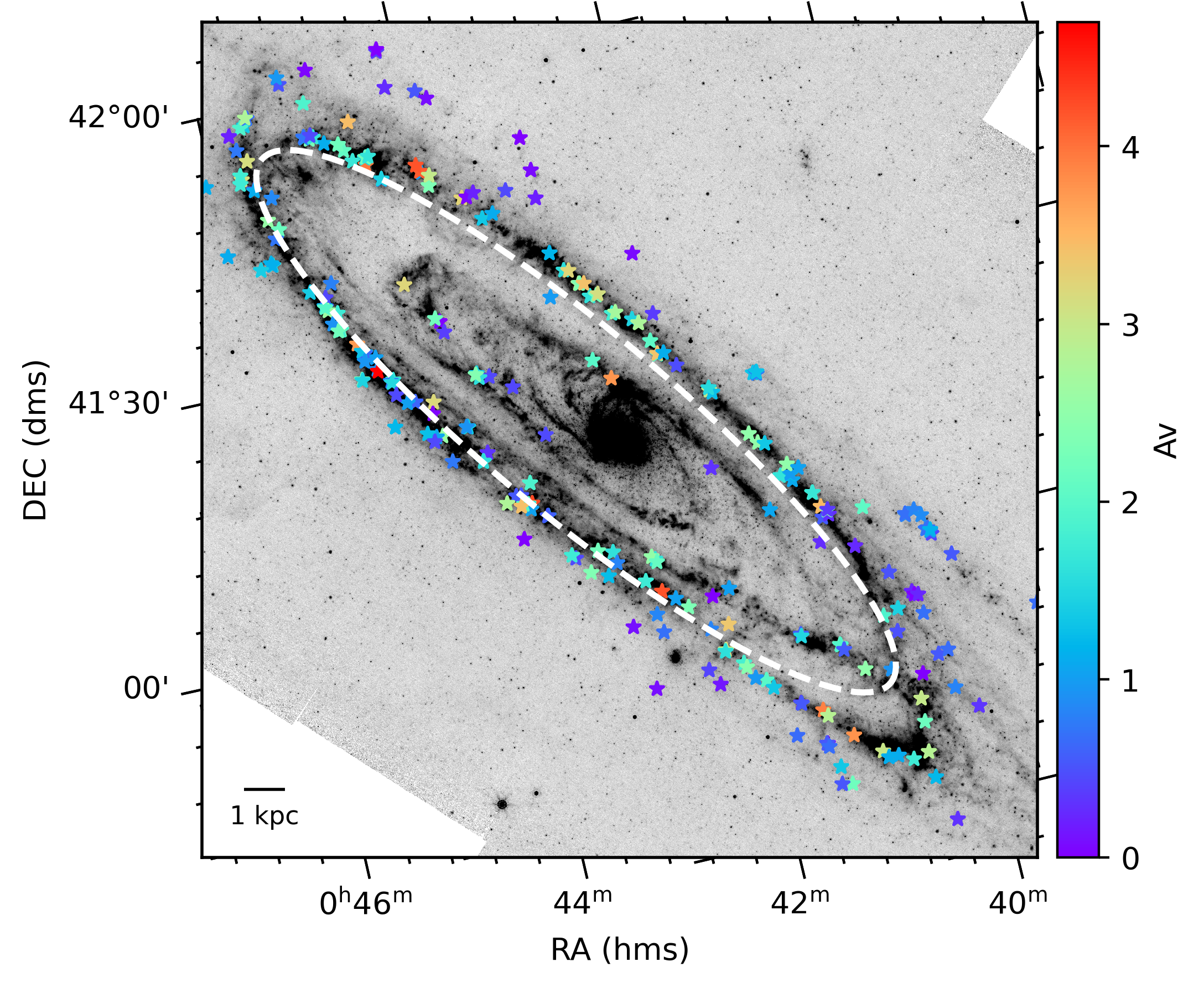}
    \caption{H~{\sc ii} regions from our sample overplotted onto the Spitzer MIPS Infrared 24 microns image \citep{Gordon_2006}, symbols are colored to values of $\rm A_{V}$. As in Figure \ref{fig:DSS}, we show the RoF fit from \citet{Gordon_2006}, in this case as a white,dashed line. The 1 kpc scalebar represents this distance on-sky at the distance of M31 (780 kpc: \citealp{Stanek_1998}).}
    \label{fig:Av}
\end{figure}

Figure \ref{fig:Av} displays the H~{\sc ii} region positions coloured by their corresponding extinction values. We find that the majority of $\rm A_{V}$ values lie within the range $\rm 0 < A_{V} < 3$. The background image, the Spitzer MIPS Infrared 24 microns image \citep{Gordon_2006}, highlights the dust distribution of M31, as expected the majority of dust is contained within the RoF. The lowest extinction values are mainly located in the outer galaxy, outside of the RoF, showing that our extinction values are consistent with the dust distribution within the galaxy. We note that there is also a range of $\rm A_{V} \sim 0 - 4.7$ for H~{\sc ii} regions within the RoF.

\subsection{Classification}
\label{sec:CLASS}

\begin{figure}
    \centering
    \includegraphics[width = 0.9\columnwidth, trim={2cm 2.2cm 2cm 2.2cm}]{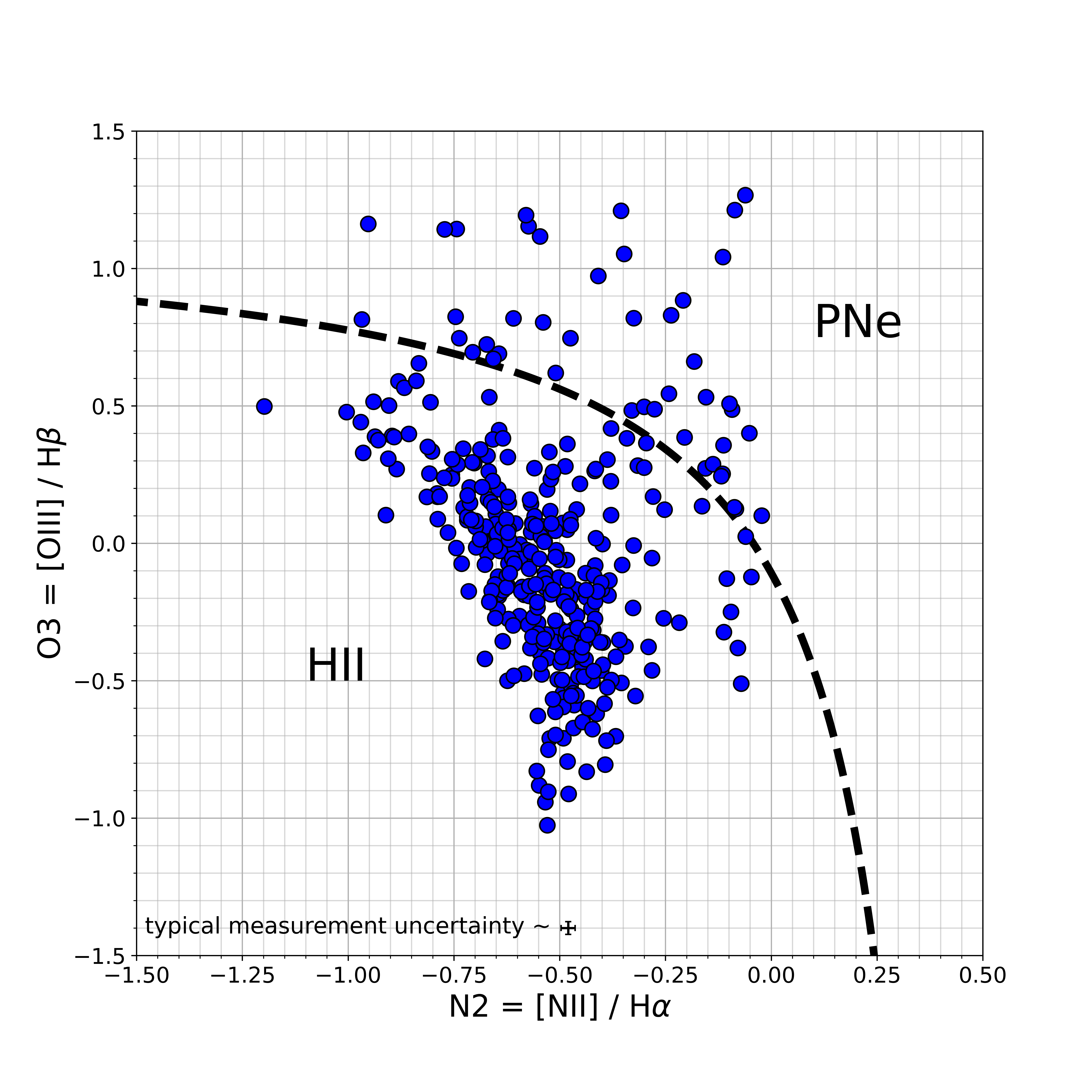}
    \caption{BPT diagram \citep{Baldwin_1981} to distinguish H~{\sc ii} regions and PNe from our optical spectra. The classification divider of \citet{Kniazev_2008} is shown. The uncertainty is derived from the mean standard deviation of repeated observations of the same sources.}
    \label{fig:BPT}
\end{figure}

Strong emission lines are a characteristic of ionized nebulae, of which H~{\sc ii} regions and PNe are two main types. Both display the same emission lines, but can be distinguished from their line ratios using a BPT (Baldwin, Phillips, Terlevich) diagram \citep{Baldwin_1981}. Line intensity ratios for H~{\sc ii} regions and PNe are different due to their primary excitation mechanisms; H~{\sc ii} regions have massive OB stars at their centres, whereas PNe have much hotter low/intermediate mass central stars. Though these can also often be distinguished by their optical morphology, classification from emission line ratios allowed us to identify even those very compact H~{\sc ii} regions that may be mistaken for PNe. We apply the classification of \cite{Kniazev_2008} using the emission line flux ratios [OIII]/H$\beta$ (O3) and [NII]/H$\alpha$ (N2). Positions of our sources on the diagram of O3 vs N2 are shown in Figure \ref{fig:BPT}. Although we ultimately classified our sources using the \cite{Kniazev_2008} classification, multiple classification dividers exist in literature (e.g. \citealp{Kewley_2001, Kauffmann_2003, Stasinska_2008}) indicating that there is an uncertainty in H~{\sc ii} region classification when a source lies close to the defined boundary. Finally, we remove another 15 sources from our H~{\sc ii} region sample which have previously been classified as Supernova Remnants (SNR) or symbiotic stars. In Figure \ref{fig:BPT}, the typical uncertainty was calculated from the mean standard deviation between repeat observations in the N2 and O3 emission line ratios.

We note that the number of H~{\sc ii} regions that we can classify from our entire dataset from the BPT diagram as shown in Figure \ref{fig:BPT} is larger than the number of H~{\sc ii} regions for which we can calculate O and N abundances (See section \ref{sec:met}). The reason for this is that the BPT diagram requires only the lines $\rm H\alpha$, $\rm H\beta$ and the stronger lines of the [OIII] and [NII] doublets ($\rm [OIII]\lambda5007$ and $\rm [NII]\lambda6584$), therefore we only require these lines to have S/N > 5. By implementing a minimum S/N for these four lines only, we obtain 416 H~{\sc ii} regions. In Figure \ref{fig:BPT} we include only those sources which have a S/N > 5 for all lines required to calculate abundances as described in Section \ref{sec:met}. Therefore, we include 309 H~{\sc ii} regions (15 of which are removed from the final sample, as previously mentioned, due to being previously classified as SNR or symbiotic stars) and 44 PNe.

\subsection{Strong-Line Diagnostics} \label{sec:met}

The use of strong-line diagnostics for our metallicity calculations allows us to probe a sample of H~{\sc ii} regions covering the entire galactic disk. In comparison, we have access to the temperature-sensitive auroral line, [OIII]$\lambda4363$, for just 4 H~{\sc ii} regions. Various methods for calculating gas-phase metallicities and abundances (mainly O and N) from strong emission lines exist in literature, derived for various different combinations of emission line ratios. Calibrations can be either empirical or theoretical (see Section 4 of \citealp{Kewley_2008} for a full review of the topic). Empirical calibrations are derived from the relationship between $\rm T_{e}$-based abundances and strong-line ratios (e.g., \citealp{PettiniPagel_2004}), whereas theoretical calibrations are derived from the relationship between strong-line ratios and a photoionization model (sometimes combined with $\rm T_{e}$-based abundances e.g., \citealp{Kewley_2002, Kobulnicky_2004}). Metallicity values are then often plotted with respect to their galactocentric radii so that a radial gradient can be calculated. \citetalias{Sanders_2012} applied five different diagnostics for O abundance to their sample of H~{\sc ii} regions in M31: \citet{Zaritsky_1994, Kewley_2002, Pilyugin_2005} and two from \citet{Nagao_2006} utilising different calibrating emission line ratios. Their results show that the diagnostic chosen affects the metallicity values calculated, along with the metallicity gradient. The uncertainty in absolute metallicity (O) values can vary up to 0.6 dex depending on the strong-line diagnostic used \citep{Kewley_2008}. 

It is for these reasons that we carefully consider our choice of strong-line diagnostic.
\defcitealias{Pilyugin_2016}{PG16}
In recent works, the strong-line abundance diagnostics of \citet{Pilyugin_2016} (hereafter \citetalias{Pilyugin_2016}) are increasingly popular for the purpose of calculating metallicities of individual H~{\sc ii} regions (e.g., \citealp{Kreckel_2019, Kreckel_2020, Williams_2022}) due to the strong agreement (within $\sim 0.1$ dex) with direct, $\rm T_{e}$ based Oxygen abundances (see \citealp{Ho_2019}). We also note that since we are observing fractions of H~{\sc ii} regions within the beam size, we are sensitive to internal ionization parameter fluctuations of H~{\sc ii} regions \citep{Jin_2023}. The calibration of \citetalias{Pilyugin_2016} provides us with two diagnostics for Oxygen abundance (R and S-calibration) and one Nitrogen abundance diagnostic (R-calibration). The following emission line ratios are utilised in the \citetalias{Pilyugin_2016} calibrations:
\begin{align*}
    \rm N_2 = (\text{[NII]}\lambda6548 + \text{[NII]}\lambda6584)/\text{H}\beta \\
    \rm R_2 = (\text{[OII]}\lambda3727 + \text{[OII]}\lambda3729)/\text{H}\beta \\
    \rm R_3 = (\text{[OIII]}\lambda4959 + \text{[OIII]}\lambda5007)/\text{H}\beta \\
    \rm S_2 = \text{[SII]}\lambda6717 + \text{[SII]}\lambda6731/\text{H}\beta.
\end{align*}
The spatial resolution of our data is not sufficient to resolve the [OII] doublet and so these lines are blended in our data. We obtain a single line flux for [OII] from a line of rest wavelength $\rm 3727 \Angstrom$ which is a sum of both lines. The R-calibration utilises the line ratios $\rm N_2, R_2$ and $\rm R_3$ whereas the S-calibration utilises the ratios $\rm N_2, S_2$ and $\rm R_3$, making these calibrations three-dimensional (3D). The high dimensionality of these diagnostics make it less sensitive to ionization parameter than 1D and 2D diagnostics. We calculate both Oxygen and Nitrogen abundance values using the \citetalias{Pilyugin_2016} method, using Nitrogen as our secondary tracer to further confirm any trends. Many strong-line diagnostics, including \citetalias{Pilyugin_2016}, separate H~{\sc ii} regions into upper (high-metallicity) and lower (low-metallicity) branches and calibrate these separately, distinguished by the $\rm N_2$ line ratio. All of our H~{\sc ii} regions lie on the upper-branch.
\defcitealias{Zaritsky_1994}{Z94}

As previously discussed, the diagnostic chosen affects the metallicity values calculated. \citetalias{Sanders_2012} have already compared four different Oxygen abundance diagnostics to their sample of H~{\sc ii} regions in M31. In order to directly compare metallicities calculated for any matching sources in our sample, we apply the O abundance diagnostic of \citet{Zaritsky_1994} (here after \citetalias{Zaritsky_1994}) also used in \citetalias{Sanders_2012}. \citetalias{Zaritsky_1994} is another empirical calibration, based on the strong-line ratio $R_{23} = (\text{[OII]}\lambda3727 + \text{[OIII]}\lambda\lambda(4959,5007))/\text{H}\beta$ \citep{Pagel_1979}. The \citetalias{Zaritsky_1994} method was chosen because it was derived from a similar dataset to ours; observations of individual H~{\sc ii} regions in spiral galaxies, including MMT observations. It is only calibrated for high-metallicity (12 + log(O/H) > 8.4 dex) H~{\sc ii} regions, which includes almost our entire sample if metallicities are calculated via the \citetalias{Zaritsky_1994} method. For the same H~{\sc ii} regions that are present in both our and the \citetalias{Sanders_2012} samples, we find that \citetalias{Zaritsky_1994}-calculated metallicities are in agreement within their respective uncertainties. The \citetalias{Zaritsky_1994} diagnostic is more sensitive to ionization parameter fluctuations than the \citetalias{Pilyugin_2016} diagnostics as it only depends on one emission line ratio, and thus we use the \citetalias{Pilyugin_2016}-derived metallicities in future discussion. O and N abundance values calculated using different calibrations for individual H~{\sc ii} regions in our sample are reported in Table \ref{tab:tab_met}

We estimate uncertainties in abundances for each diagnostic measured, again using the standard deviation of repeat observations of the same source, reflecting the total empirical error budget. Where a source has just one corresponding observation, the quoted uncertainty in Table \ref{tab:tab_met} is the mean percentage uncertainty calculated for sources for which we have multiple observations. Additionally, we introduce a lower bound of 0.01 to uncertainties, again to reflect unquantifiable systematic uncertainties. The standard deviation probes the overall error budget accounting for differences in atmospheric conditions between observations and slight differences between the individual Hectospec optical fibres. Uncertainties in absolute abundance due to the chosen metallicity diagnostic may be as large as 0.7 dex \citep{Kewley_2008} (although empirical calibrations produce O abundances that are in better agreement with directly measured abundances), however, we are interested in metallicity differences between individual H~{\sc ii} regions more than the absolute metallicity values. Work analysing the reliability of relative H~{\sc ii} region O abundances have shown that these are statistically robust between different calibrations (e.g., \citealp{Ho_2017}). Finally, we note that selection effects may occur due to emission line detection; any spectra without all of the required lines detected at sufficient S/N could not be included in analysis. However, since we are analysing strong emission lines, this likely has little affect on resulting trends. 

\begin{table*}
    \centering
    \caption{Chemical abundances for M31 H~{\sc ii} regions. Nominal uncertainties are estimated from the standard deviation between values obtained from repeat observations of the same source, where available (see text), with a lower bound of 0.01.}
    \label{tab:tab_met}
    \begin{tabular*}{\linewidth}{@{\extracolsep{\fill}} ccccccccccccc}
        \hline
         ID & R.A., Dec. & R $\rm ^{a}$ & Av$\rm ^{b}$ & (O/H)* R-cal$\rm ^{c}$ & (O/H)* S-cal$\rm ^{d}$ &  (O/H)* Z94$\rm ^{e}$ & (N/H)* R-cal$\rm ^{f}$ \\ & (J2000) & (kpc) & (mag) \\
         \hline
         1 & 0:39:13.01, 40:41:44.88 & 16.07 & 1.00 $\pm$ 0.13 & 8.42 $\pm$ 0.02 & 8.50 $\pm$ 0.01 & 8.89 $\pm$ 0.03 & 7.34 $\pm$ 0.05 \\ 2 & 0:39:14.70, 40:48:33.12 & 18.33 & 0.66 $\pm$ 0.15 & 8.36 $\pm$ 0.05 & 8.47 $\pm$ 0.03 & 8.71 $\pm$ 0.18 & 7.16 $\pm$ 0.15 \\ 3 & 0:39:16.50, 40:41:04.92 & 15.44 & 0.08 $\pm$ 0.04 & 8.39 $\pm$ 0.02 & 8.47 $\pm$ 0.01 & 8.69 $\pm$ 0.01 & 7.26 $\pm$ 0.03 \\ 4 & 0:39:55.39, 40:55:49.18 & 15.56 & 0.54 $\pm$ 0.01 & 8.55 $\pm$ 0.01 & 8.61 $\pm$ 0.01 & 9.01 $\pm$ 0.05 & 7.67 $\pm$ 0.05 \\ 5 & 0:40:00.60, 40:39:12.60 & 11.00 & 0.33 $\pm$ 0.08 & 8.57 $\pm$ 0.01 & 8.67 $\pm$ 0.01 & 8.90 $\pm$ 0.03 & 7.64 $\pm$ 0.04 \\ 6 & 0:40:03.89, 40:58:27.12 & 15.51 & 0.40 $\pm$ 0.15 & 8.62 $\pm$ 0.01 & 8.58 $\pm$ 0.02 & 8.75 $\pm$ 0.05 & 7.75 $\pm$ 0.04 \\ 7 & 0:40:04.01, 40:58:53.40 & 15.69 & 1.16 $\pm$ 0.01 & 8.44 $\pm$ 0.01 & 8.49 $\pm$ 0.01 & 8.90 $\pm$ 0.01 & 7.42 $\pm$ 0.03 \\8 & 0:40:04.30, 40:58:45.84 & 15.59 & 0.88 $\pm$ 0.08 & 8.61 $\pm$ 0.02 & 8.58 $\pm$ 0.01 & 8.60 $\pm$ 0.12 & 7.69 $\pm$ 0.09 \\ 9 & 0:40:05.30, 40:59:07.44 & 15.61 & 0.68 $\pm$ 0.14 & 8.59 $\pm$ 0.02 & 8.55 $\pm$ 0.01 & 8.66 $\pm$ 0.11 & 7.67 $\pm$ 0.11 \\ 10 & 0:40:06.36, 40:59:04.05 & 15.43 & 0.75 $\pm$ 0.13 & 8.50 $\pm$ 0.02 & 8.45 $\pm$ 0.01 & 8.84 $\pm$ 0.03 & 7.61 $\pm$ 0.05 \\
         \hline
    \end{tabular*}
\begin{minipage}{\linewidth}~\\
    $\rm ^{a}$ Source galactocentric radius, calculated as described in Section \ref{sec:results}. $\rm ^{b}$ Extinction values calculated as described in Section \ref{sec:EC}. $\rm ^{c}$ Oxygen abundance values calculated using the \citetalias{Pilyugin_2016} R-calibration. $\rm ^{d}$ Oxygen abundance values calculated using the \citetalias{Pilyugin_2016} S-calibration.  $\rm ^{e}$ Oxygen abundance values calculated using the \citetalias{Zaritsky_1994} calibration.  $\rm ^{f}$ Nitrogen abundance values calculated using the \citetalias{Pilyugin_2016} R-calibration. The full table of 294 sources is available online, we show a portion here for 10 sources. * (O/H) = 12 + log(O/H) and (N/H) = 12 + log(N/H).
\end{minipage}
\end{table*}

\section{Results and Discussion} \label{sec:results}

We derived the radial O and N abundance gradients across M31 for H~{\sc ii} regions using strong-line methods \citetalias{Zaritsky_1994} and \citetalias{Pilyugin_2016} as outlined in section \ref{sec:met}. The Galactocentric radius of each H~{\sc ii} region was calculated following \cite{Haud_1981}. We assume the galactic centre of M31 and its position angle (PA) to be: 
\begin{align*}
\rm \alpha_{0} = 00^{\circ}42^{\prime}44^{\prime\prime}.52~~(J2000) \\
\delta_{0} = +41^{\circ}16^{\prime}08^{\prime\prime}.69)~~(J2000) \\
\Theta_{0} = 37^{\circ}42^{\prime}54^{\prime\prime}
\end{align*}
\citep{Haud_1981} and the angle of inclination to be $i = 12.5^{\circ}$ \citep{Simien_1979}. We note that these deprojected distances from the Galactic centre of M31 rely on the assumption that all positions are within the same plane, and so some separations may be underestimated. 

The following sub-sections are structured as follows. First, we present results for radial O and N abundance gradients in M31. Then, we subtract the radial trend from our data in order to analyse reduced and enhanced abundances (compared to the radial average) and search for azimuthal trends around the RoF or variation with spiral arm structure. We look at the two-point correlation function of metallicity with source separation scale and compare to other galaxies from \cite{Kreckel_2020} and \cite{Williams_2022}. Finally, we analyse the relationship between H~{\sc ii} region metallicities and the conversion factor for $\rm ^{12}CO$ luminosity to dust mass ($\rm M_{dust}$), $\alpha^{\prime}(^{12}{\rm CO})$. 

\subsection{Radial Abundance Gradients} \label{sec: grads}

\begin{figure*}
    \centering
    \subfigure{
        \includegraphics[width=0.48\textwidth, trim={3.5cm 2cm 2.5cm 2cm}]{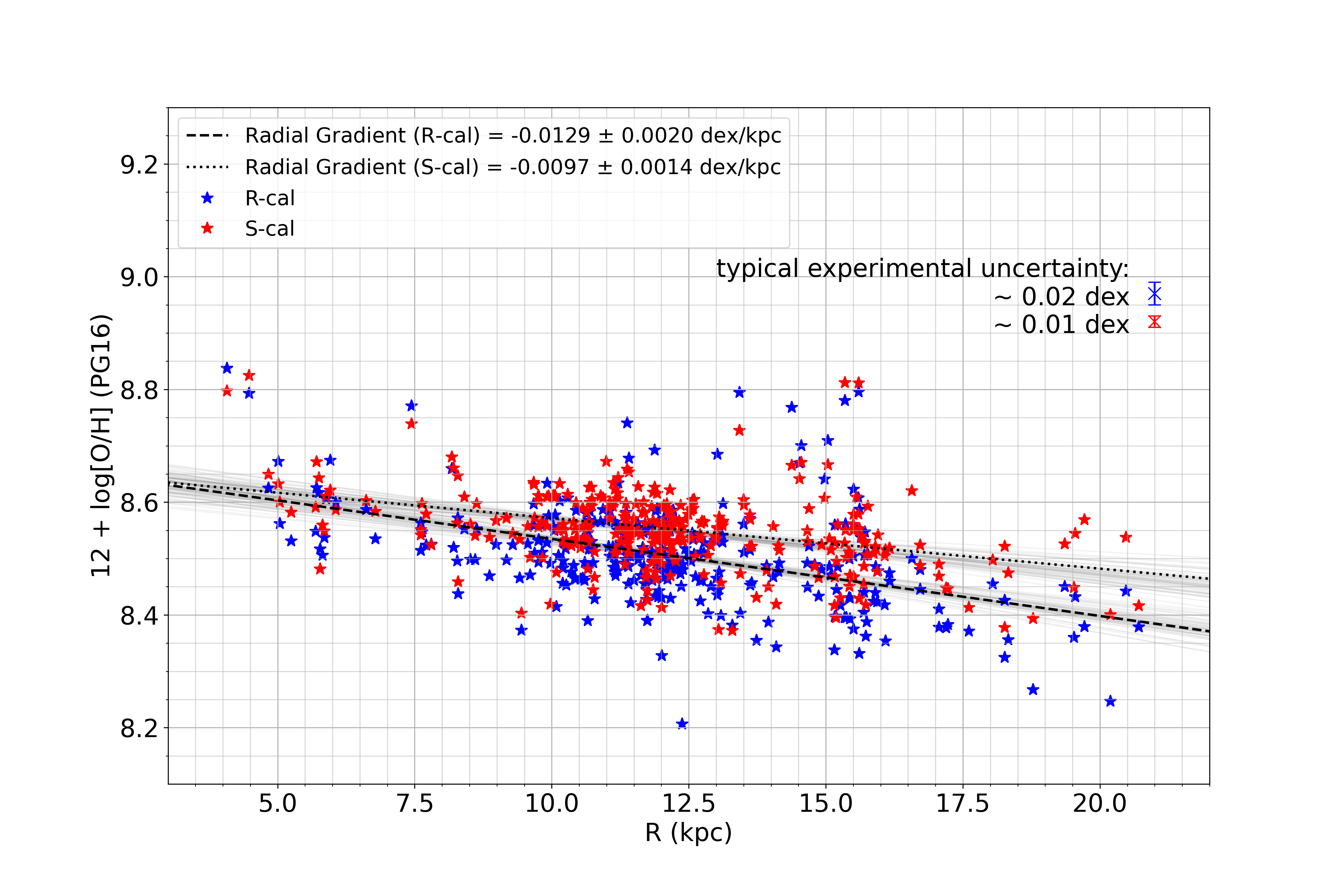}}
    \hspace{0.0\textwidth} 
    \subfigure{
        \includegraphics[width=0.48\textwidth, trim={2.5cm 2cm 3.5cm 2cm}]{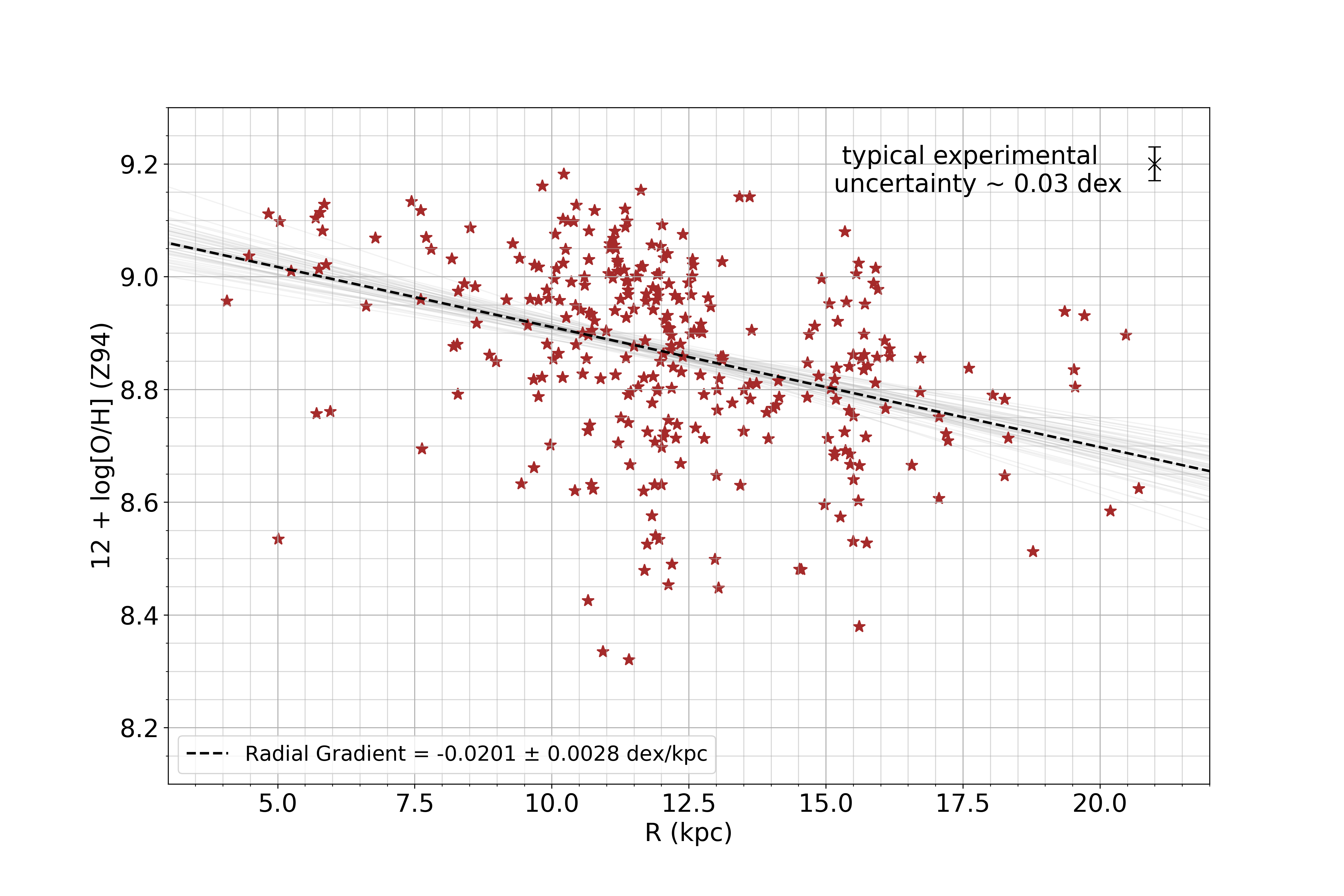}}
    \caption{Oxygen abundance (12 + log[O/H]) for H~{\sc ii} regions in M31 from the LEFT: \citetalias{Pilyugin_2016} R and S-calibrations RIGHT: \citetalias{Zaritsky_1994} diagnostics as a function of galactocentric radius (R kpc).}
    \label{fig:Ograds}
\end{figure*}

Figure \ref{fig:Ograds} shows H~{\sc ii} region metallicities as a function of GCR. We show the typical uncertainty of a given measurement as the mean standard deviation between repeat observations of the same sources. We then calculated the radial metallicity (O abundance) gradients for our sample of 294 H~{\sc ii} regions from each calibrations, calculated from bootstrap resampling of least-squares fitting. Gradient uncertainties correspond to the standard deviation of the individual bootstrap gradient calculations.

For the \citetalias{Pilyugin_2016} calibrations we find radial gradients of $-0.0129 \pm 0.0020 \text{ dex kpc}^{-1}$ (R-calibration) and $-0.0097 \pm 0.0014 \text{ dex kpc}^{-1}$ (S-calibration), which are in agreement within $2\sigma$. Both values are in agreement within $1\sigma$ with values that \citetalias{Sanders_2012} obtained using the diagnostics of \citet{Kewley_2002} and \citet{Nagao_2006}. The S-calibration result is also consistent with the \citetalias{Sanders_2012} result using the \citet{Pilyugin_2005} diagnostic within $1\sigma$. Thus, despite improving on individual metallicity uncertainties due to ionization parameter fluctuation, the radial metallicity gradient derived from \citetalias{Pilyugin_2016}-derived metallicities is still somewhat dependent on the calibration method. Separating the galaxy into two halves down the minor axis, we calculated radial gradients for the two halves of the galaxy, obtaining $-0.0103 \pm 0.0022 \text{ dex kpc}^{-1}$ for $\rm PA < 180\deg$ and $-0.0066 \pm 0.0024 \text{ dex kpc}^{-1}$ for $\rm PA > 180\deg$. The values are both consistent with the radial gradient of the full sample and with one another within $\rm 1 \sigma$ uncertainties. 

The metallicity (O abundance) gradient we calculate from the \citetalias{Zaritsky_1994} diagnostic as shown in Figure \ref{fig:Ograds} (RIGHT) is $-0.0201 \pm 0.0028 \text{ dex kpc}^{-1}$. This result is in agreement within $1\sigma$ with the radial metallicity gradient calculated by \cite{Sanders_2012} also using \citetalias{Zaritsky_1994}-derived abundances ($-0.0208 \pm 0.0048 \text{ dex kpc}^{-1}$) and that calculated by \citetalias{Zaritsky_1994} ($-0.020 \pm 0.007 \text{ dex kpc}^{-1}$). The radial metallicity gradients in our work calculated from the \citetalias{Zaritsky_1994} and the \citetalias{Pilyugin_2016} R-calibration agree within $2\sigma$ (\citetalias{Zaritsky_1994} and \citetalias{Pilyugin_2016} S-calibration agree within $3\sigma$). Therefore the radial metallicity gradient of M31 is again dependent on which strong-line diagnostic is chosen to calculate individual H~{\sc ii} region metallicities.

\citet{Wenger_2019} built on previous Milky Way (MW) H~{\sc ii} region studies (e.g., \citealp{Balser_2011}) finding an O abundance gradient of $-0.052 \pm 0.004 \text{ dex kpc}^{-1}$ from $T_e$ line measurements. A radial O abundance gradient of $-0.041 \pm 0.006 \text{ dex kpc}^{-1}$ was calculated by \citet{Esteban_2018}. Compared to the O abundance gradients calculated in this work from \citetalias{Pilyugin_2016} diagnostics (which has agreement within $\sim$ 0.1 dex with direct, $T_{e}$-based metallicity measurements; \citealp{Ho_2019}), the radial gradient of the MW is $\sim$ 4 - 6 times steeper than that of M31. \citet{Kreckel_2019} found O abundance gradient values for their sample of 8 spiral galaxies, also derived from \citetalias{Pilyugin_2016} S-calibration abundances, ranging from -0.0199 to $-0.1586 \text{ dex kpc}^{-1}$ (excluding one galaxy which produced a positive gradient). Therefore the O abundance gradient of M31 is shallow compared to these spiral galaxies. 

\begin{figure}
    \centering
    \includegraphics[width = \columnwidth, trim={2.5cm 2cm 2.5cm 2cm}]{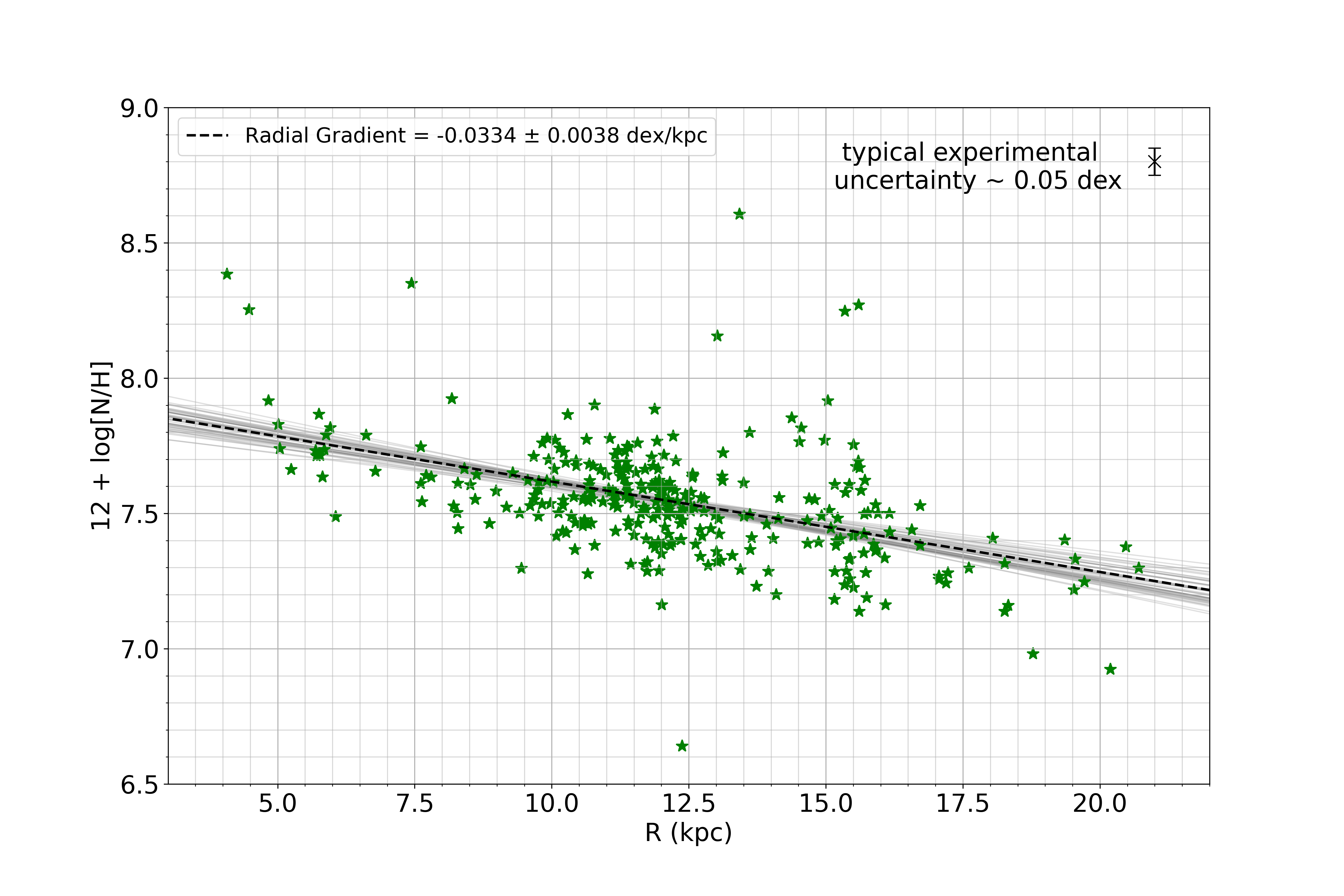}
    \caption{Nitrogen abundance (12 + log[N/H]) for H~{\sc ii} regions in M31 from the \citetalias{Pilyugin_2016} R-calibration as a function of galactocentric radius (R kpc).} 
    \label{fig:PG16N}
\end{figure}

Again from bootstrap resampling, we calculate an N abundance gradient of $-0.0334 \pm 0.0038 \text{ dex kpc}^{-1}$ from 294 H~{\sc ii} regions using the R-calibration of \citetalias{Pilyugin_2016}. We display this result in Figure \ref{fig:PG16N}. The N abundance gradient is significantly larger than the O abundance gradient by a factor of $\sim$ 1.7 - 3.4 depending on the choice of strong-line diagnostic. This is consistent with simulations from \cite{Vincenzo_2018} where they find increasing N/O with O/H in spatially resolved star-forming disc galaxies, and a resulting negative radial N/O abundance gradient.

The N abundance gradient calculated in this work is compatible with that from \citetalias{Sanders_2012} ($- 0.0303 \pm 0.0049$) which uses the strong line diagnostic of \citet{Pilyugin_2010}, an older version of the N abundance diagnostic from \citetalias{Pilyugin_2016}. For the MW, \citet{Esteban_2018} found N abundance gradients ranging between $-0.047 \text{ to} - 0.050 \pm 0.008 \text{ dex kpc}^{-1}$ from direct H~{\sc ii} region abundances of 13 sources. This is steeper than the N abundance gradient found in this work for M31 by a factor of $\sim$ 1.4 - 1.5. As found for the O abundance gradient, there is statistically significant scatter around the N abundance gradient, thus providing further evidence that there is abundance variation beyond the radial gradient. Figures \ref{fig:Ograds} \& \ref{fig:PG16N} show that H~{\sc ii} regions at a similar Galactocentric radius can have very different metallicities, as also shown in \citetalias{Sanders_2012} in their Figures 8 \& 14.

Although our abundance gradients are in agreement with previous work, Figures \ref{fig:Ograds} \& \ref{fig:PG16N} show that there is significant (greater than the typical uncertainty) scatter around the radial gradient at all Galactocentric radii. The majority of M31 H~{\sc ii} regions are located at $\sim$ 12 and 15 kpc. These correspond to the approximate locations of the main RoF and its outer component at \citep{Lewis_2015}. If the metallicity variation throughout M31 could be fully explained by just the radial gradient, we would expect that these sources all have a similar metallicity. However, in Figure \ref{fig:Ograds} (LEFT) we see O abundances ranging from $\sim8.35 - 8.75$ dex (excluding two outliers), from \citetalias{Pilyugin_2016} S-calibration-based abundances, for sources within the RoF ($\sim 10 - 16$ kpc). The metallicity range covered by RoF H~{\sc ii} regions is even larger for the \citetalias{Zaritsky_1994} diagnostic ($\sim8.4 - 9.2$ dex) though this is partially due to the inherent noise of the diagnostic. \citetalias{Sanders_2012} found no dependence of metallicity or gradient on the surface brightness of H~{\sc ii} regions for their similar sample. 

Our final H~{\sc ii} region sample requires all emission lines required for the strong line calibrations (see Section \ref{sec:met}) to have S/N > 5, producing a sample size is reduced to 294. We can, theoretically, extend our sample size by assuming a 3:1 ratio for the strong to weak lines in the [OIII] and [NII] doublets \citep{Storey_2000}. We investigated this following the method outlined in \citet{Kreckel_2020}: we measure only the lines $\text{[OIII]}\lambda5007$ and $\text{[NII]}\lambda6584$, the stronger lines in each doublet, and divide these line fluxes by 3 to obtain the line fluxes of $\text{[OIII]}\lambda4859$ and $\text{[NII]}\lambda6548$. Therefore, the condition that the weaker lines of the doublet must have S/N > 5 for the abundances to be recorded can be eliminated, increasing the number of H~{\sc ii} regions to 322 . This is lower than the 416 classified H~{\sc ii} regions (see Section \ref{sec:CLASS}) because both of the lines in the [SII] doublet are still required to have S/N > 5. Calculating the radial metallicity gradients for the extended sample, we obtain values compatible within $1\sigma$ to the previously reported values in this section, for all abundance diagnostics and tracers. Ultimately, the 3:1 ratio assumption does not significantly increase our sample size and leads to no change to our results. Therefore we use our sample of 294 H~{\sc ii} regions, measuring both lines in the [OIII] and [NII] doublets, for all further analysis.

\subsection{Non-radial trends} \label{sec:nonrad}

\begin{figure}
    \centering
    \includegraphics[width = \columnwidth, trim={3cm 3cm 3cm 3cm}]{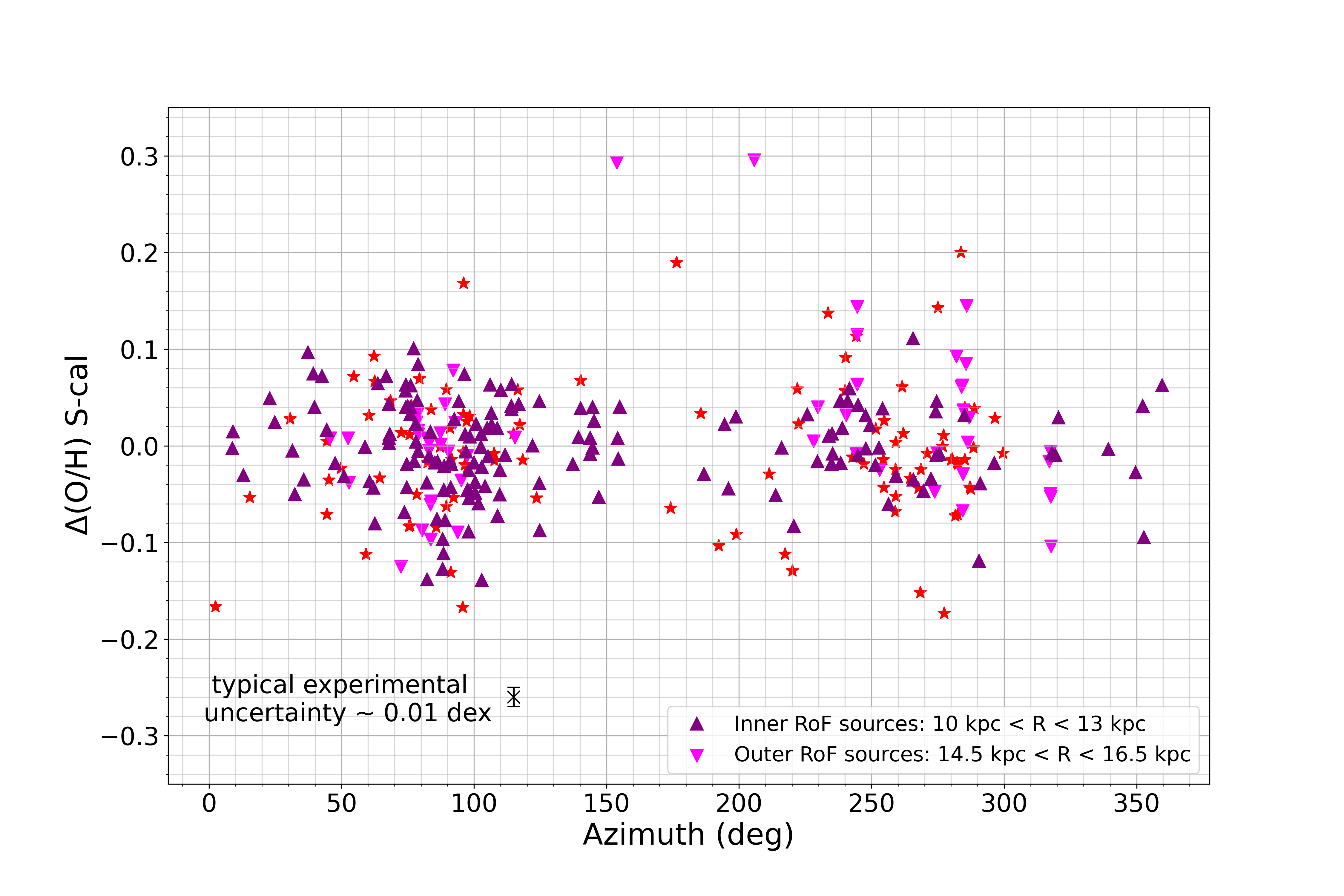}
    \caption{Oxygen abundance (12 + log[O/H]) for H~{\sc ii} regions in M31 from the \citetalias{Pilyugin_2016} S-calibration as a function of azimuthal angle with 0 deg corresponding to the minor axis in the northern direction and moving in the clockwise direction. Sources which are located within the main RoF component at $\sim 10 - 13 \rm~kpc$ are shown in dark purple. Sources located within the outer RoF component at $\sim 14.5 - 16.5 \rm~kpc$ are shown in pink.}
    \label{fig:Az}
\end{figure}

We next subtract the S-cal radial-gradient from individual H~{\sc ii} region Oxygen abundance values, so that only residual metallicity remains. We chose to use the S-cal abundance values for this purpose because the [SII] lines are less affected by extinction than the [OII] line (see Fig. 3 from \citealp{Cardelli_1989}), decreasing the inherent uncertainty. This is also identical to the calibration used in other works investigating H~{\sc ii} region metallicity scatter across galaxy disks (e.g., \citealp{Kreckel_2019}) enabling us to make a direct comparison without the uncertainty due to using different calibrations. This enables us to investigate any second-order trends such as azimuthal effects and variations corresponding to galactic features such as spiral arms and the RoF. In Figure \ref{fig:Az} we show $\Delta$(O/H) of H~{\sc ii} regions as a function of azimuthal position in the galaxy for the \citetalias{Pilyugin_2016} S-calibration. From this plot we find little evidence of systematic metallicity trends in the azimuthal direction, but that scatter of standard deviation 0.06 dex around the radial gradient occurs across the entire galaxy. This is true even for sources at a similar galactocentric radius, as shown for the sources located within the RoF at $\sim 10 - 13 \rm~kpc$ as well as its outer component at $\sim 14.5 - 16.5 \rm~kpc$, highlighted on Figure \ref{fig:Az}. 

Previous studies have shown that significant scatter around the radial gradient typically occurs distributed across the galaxy \citep{Kreckel_2019, Grasha_2022}. Alongside this, studies have shown hints of distinct azimuthal variation in the form of enhanced abundances along spiral arms (e.g., \citealp{Kreckel_2019, Grasha_2022}), although statistical tests failed to quantify a systematic correlation due to the large scatter present across the entire galaxy. On the other hand, \citet{Ho_2017} found a statistical difference between spiral-arm and inter-arm regions, finding that 76\% of H~{\sc ii} regions within the main spiral are above the radial gradient whilst only 19\% of inter-arm regions are above the gradient. We note that in these spiral galaxies, the majority of H~{\sc ii} regions are located within the spiral arms, however, in M31 the majority are located in the RoF. For H~{\sc ii} regions located in the RoF, we see an approx. equal number of reduced abundance H~{\sc ii} regions as we do enhanced abundance H~{\sc ii} regions. This could suggest that we are seeing the effect of a recent collision with M32 adding some more pristine gas that would otherwise not be present, predicted to have occured $\sim 200 - 800$ Myr ago.

A strong positive correlation between $\Delta$(O/H) and ionization parameter hints that local enrichment is linked to differences in the local physical conditions (\citealp{Grasha_2022}). Along with evidence that H~{\sc ii} regions with enhanced abundances have high ionization parameter, H$\alpha$ luminosity, younger star clusters, and high CO in the associated molecular clouds \citep{Kreckel_2019}, this suggests that local enrichment has occurred due to recent star formation. \citet{Kreckel_2019} see correlation between reduced abundance H~{\sc ii} regions and an increased H$\alpha$ velocity dispersion which indicates the dynamical state of the ionized gas. This indicates that large-scale mixing may have taken place, introducing pristine material from outside the galaxy.

\begin{figure}
    \centering
    \includegraphics[width = \columnwidth, trim={3cm 3cm 3cm 3cm}]{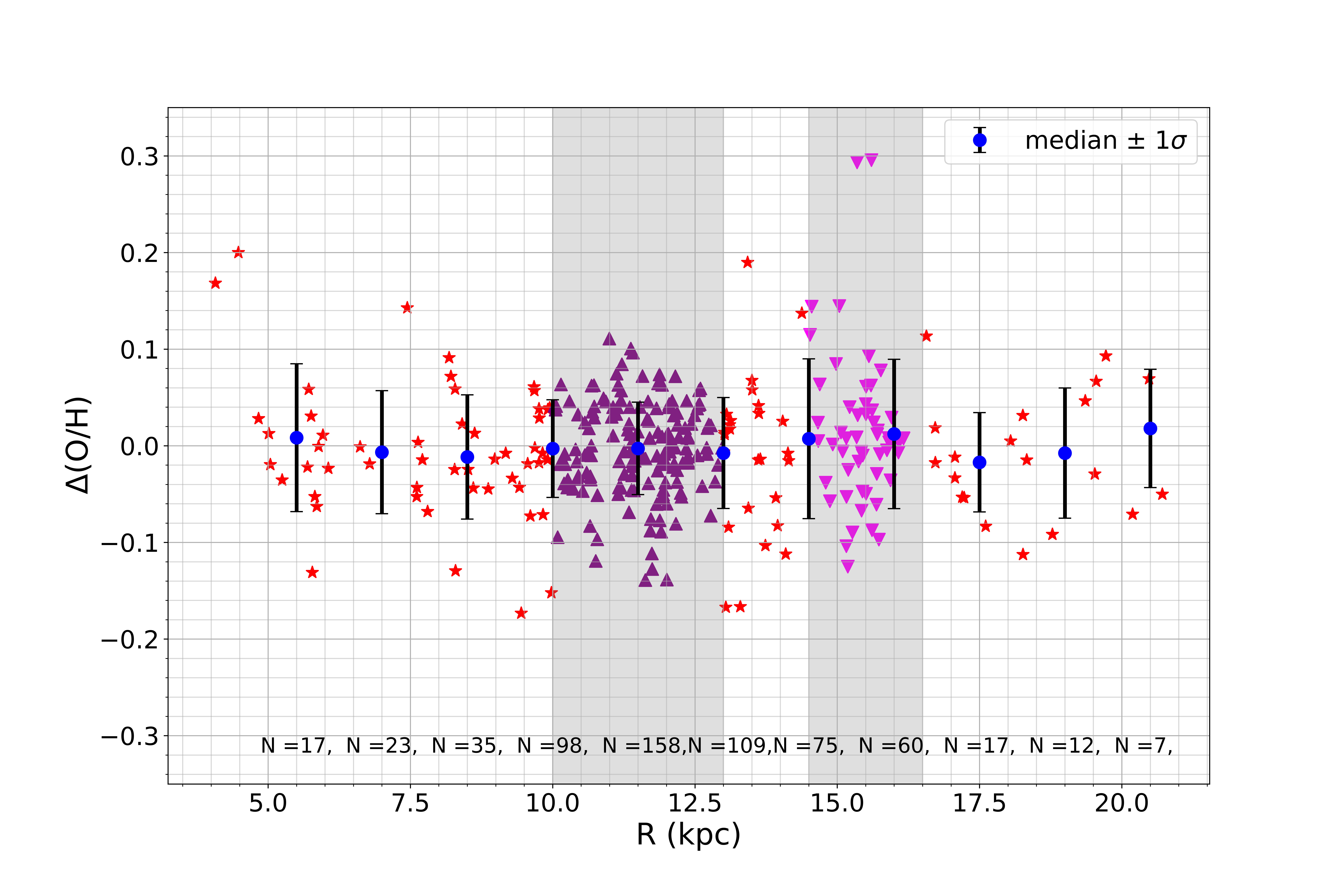}
    \caption{Variation of $\Delta$(O/H) (\citetalias{Pilyugin_2016} S-calibration) with galactocentric radius, with individual H~{\sc ii} regions shown in red and the mean values for bins of equal radial size (3 kpc) shown as blue circles. Bins are separated by 1.5 kpc and overlap by 50\%. Sources located within the main RoF at $\sim 10 - 13 \rm kpc$ are shown in dark purple and sources located within the outer RoF component at $\sim 14.5 - 16.5 \rm kpc$ are shown in pink. In grey we show the approximate positions and widths of the RoF components, based on where the majority of our sources lie. In this case, the black bars show the standard deviation dispersion of the sources in each bin, which have N number of sources.}
    \label{fig:EqRBins}
\end{figure}

\begin{figure}
    \centering
    \includegraphics[width = \columnwidth, trim={3cm 4cm 3cm 4.7cm}]{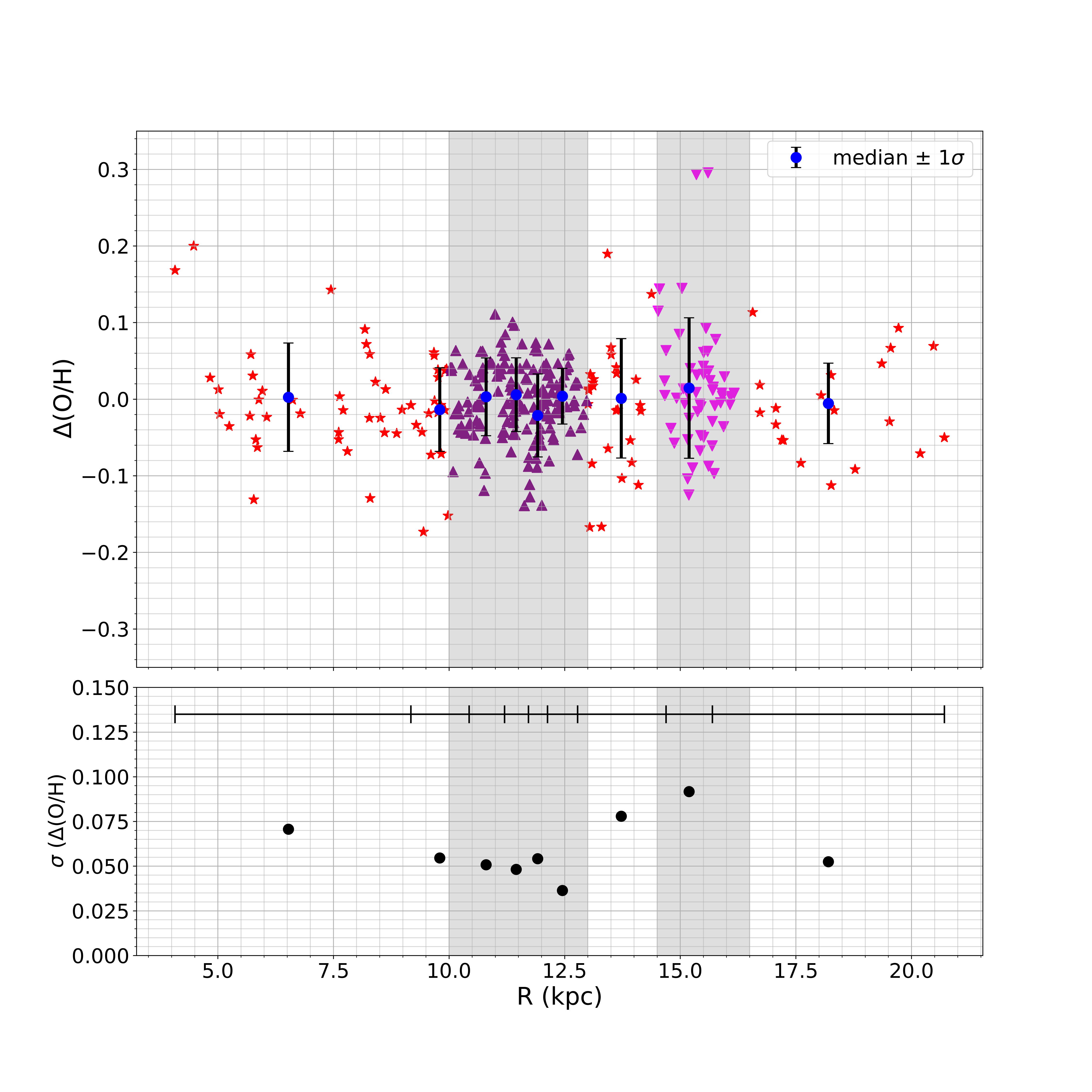}
\caption{TOP: The same as Figure \ref{fig:EqRBins} but for bins of approx. equal number of sources (N = 34 or 35). Sources located within the RoF at $\sim 10 - 13 \rm kpc$ are shown in dark purple and sources located within the outer RoF component at $\sim 14.5 - 16.5 \rm kpc$ are shown in pink. The grey regions represent approximate RoF positions, the same as for Figure \ref{fig:EqRBins}. Vertical black bars give the standard deviation of sources in each bin. BOTTOM: Standard deviation with varying R. The black bars show the radial range covered by each bin, relevant to both figures.}
\label{fig:EqNBins}
\end{figure}

We suggest that a similar explanation may be relevant to M31. In Figures \ref{fig:EqRBins} \& \ref{fig:EqNBins} we show the mean and standard deviation $\Delta$(O/H) at varying bins of galactocentric radius. Figure \ref{fig:EqRBins} uses bins of equal radial widths (3 kpc) that overlap by 50\% of the bin size. For Figure \ref{fig:EqNBins} we use bins containing approximately equal numbers of sources, where N = 34 or 35. In both figures we see that the standard deviation in the scatter of $\Delta$(O/H) is of the range $\sim$ 0.03 - 0.10 dex and that approx. $\sim$ 30\% of H~{\sc ii} regions have |$\Delta$(O/H)| > $1\sigma$ and 20\% of those outliers have |$\Delta$(O/H)| > $2\sigma$. Approximately 50\% of outliers are enhanced abundances (and approximately 50\% are reduced abundances) compared to the metallicity gradient. The greatest amount of scatter occurs at R $\sim$ 15 kpc which corresponds with the outer RoF (the corresponding sources are shown as pink upside-down triangles), this is also where we see the most enhanced abundance H~{\sc ii} regions. In comparison, the main RoF at $\sim 10 - 13 \rm kpc$ is where we see the least scatter, suggesting that this material is more well-mixed than in the outer RoF component. However, as we see both enhanced and reduced abundances of |$\Delta$(O/H)| > $1\sigma$ distributed across the entire disk, general systematic trends have not been identified. We note that we make no statement on the exact location and size of the RoF, and our approximation is based on the distribution of our H~{\sc ii} region sample.

The presence of enhanced abundance H~{\sc ii} regions indicates that material has likely been enriched due to recent star formation and supernovae (SNe). Simulations indicate that kpc-scale mixing can take 100 - 350 Myr \citep{Roy_1995, deAvillez_2002}. Simulations performed by \citet{deAvillez_2002} indicate a mixing timescale of 350 Myr for the Galactic SN rate, therefore it is possible that there is SN-enriched material in M31, primarily within the RoF, that is not yet fully mixed, causing inhomogenities in the ISM to remain. 

A possible explanation for the pristine gas is that it may have been brought into M31 from outside of the galaxy, resulting in reduced abundance H~{\sc ii} regions. A possible source of this pristine gas is a recent collision with M32 which models predict occurred $\sim 200 - 800$ Myr ago (\citealp{Block_2006, Davidge_2012, Dierickx_2014}). Lower-metallicity (in comparison to the metallicity of gas in M31) gas may have been stripped from M32 and not yet fully mixed into M31, especially if the collision occurred closer to $\sim 200$ Myr ago. Most $1\sigma$ outliers are located within the RoF, which is believed to have formed during the collision with M32 \citep{Braun_1991}. The differences in the standard deviation of the residual scatter are also displayed in Figure \ref{fig:EqNBins}. The global scatter over all values of R is $\sim 0.06$ dex. However, we see in Figure \ref{fig:EqNBins} that the standard deviation can be as low as <  0.04 dex within the main/inner RoF, suggesting that this material is well mixed. On the contrary, the standard deviation scatter is as high as > 0.09 dex for the outer RoF. In Figures \ref{fig:EqRBins} \& \ref{fig:EqNBins} we see increased scatter in $\Delta$(O/H) at R $\sim$ 15 kpc, suggesting that the material in the outer RoF is less well-mixed than the inner RoF. For comparison, \citet{Kreckel_2019} found global scatter of 0.03 - 0.05 dex for their galaxies. 

\subsection{Two-Point Correlation Function of Metallicity} \label{sec:local}

\begin{figure*}
\centering
\includegraphics[width=\textwidth,trim={2cm 1cm 2cm 1cm}]{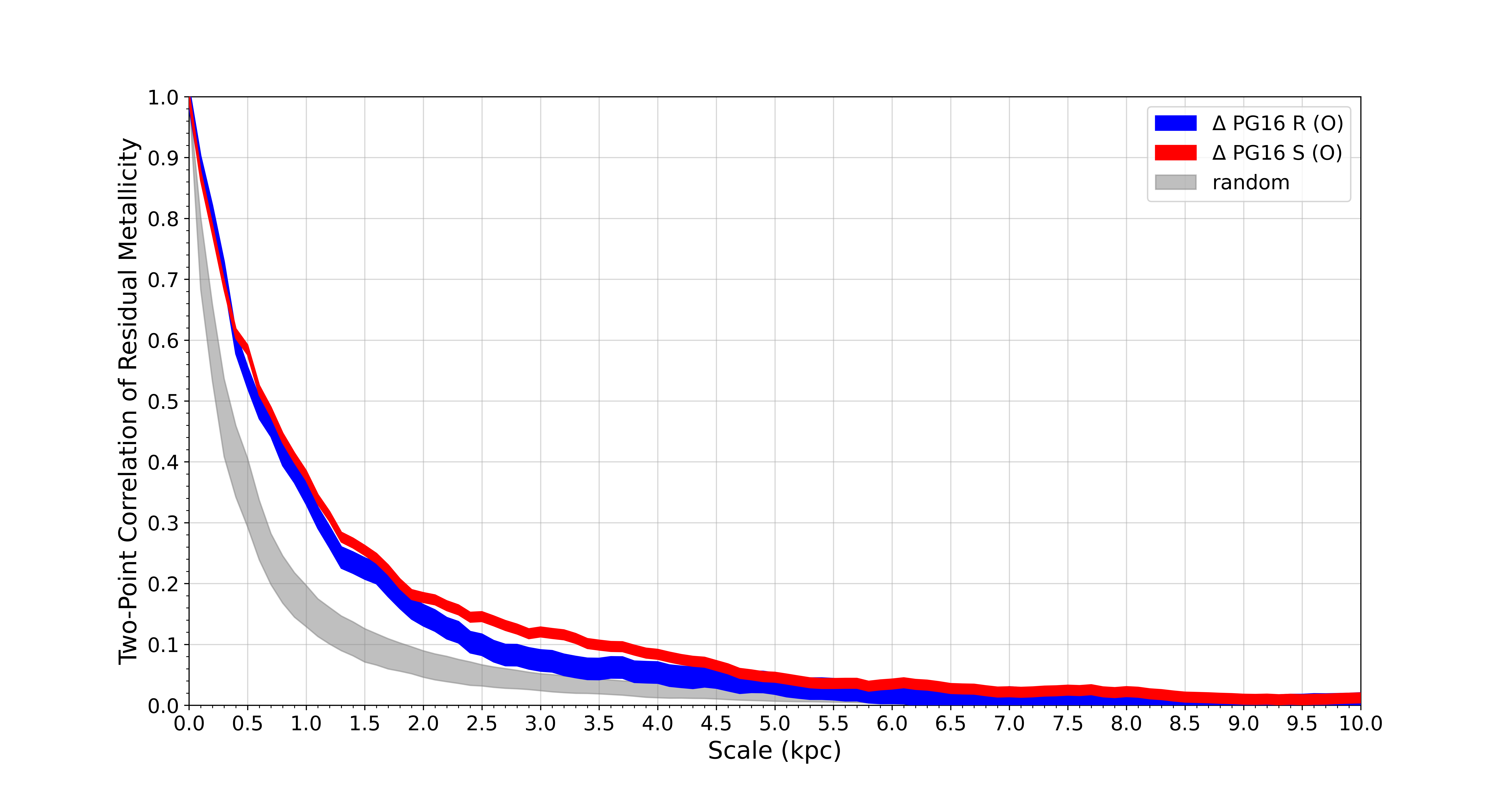}
\caption{Two-point correlation of metallicity for the \citetalias{Pilyugin_2016} strong-line diagnostics as a function of spatial scale (H~{\sc ii} regions separated by up to 10 kpc). In grey, we show the two-point correlation if the metallicity values are randomised i.e. metallicity and position within the galaxy are uncorrelated.}
\label{fig:2p}
\end{figure*}

Following the method of \citet{Kreckel_2020}, we calculate the two-point correlation function of residual metallicity as a function of H~{\sc ii} region separation after the radial trend has been subtracted (Figure \ref{fig:2p}) using their equation 5. As was also done in their work, we computed a randomized sample by randomly shuffling the metallicity values amongst our H~{\sc ii} region positions. For our purpose we used metallicity (O) values from the \citetalias{Pilyugin_2016} S-calibration. Then we calculate the correlation function via bootstrap resampling and taking the average and standard deviation over 100 iterations. From this we investigate the separation between H~{\sc ii} regions where we no longer see the same degree of homogeneity above that expected from the randomised sample. 

Our two-point correlation function is shown in Figure \ref{fig:2p}. We show that the correlation falls below 0.5 (the 50\% correlation scale) at $\sim0.6 $ kpc separation and below 0.3 at $\sim1.2 $ kpc separation. This is significantly different to the randomised sample for which the correlation falls below 0.5 at $\sim0.3 $ kpc separation and below 0.3 at $\sim0.6 $ kpc separation. From this we conclude that there is correlation in H~{\sc ii} region metallicities for nearby sources above what we would see if the scatter around the radial gradient was completely random. The two-point correlation functions for the 19 PHANGS-MUSE spiral galaxies are reported in \cite{Williams_2022}. The 50\% correlation scale values found mostly range from 0.19 - 1.08 kpc apart from for one galaxy which reaches 0.5 correlation at 4.12 kpc. This suggests that M31 is quite well-correlated on small scales which is similar in general to these galaxies, when comparisons are made using the same strong-line diagnostic (\citetalias{Pilyugin_2016} S-calibration). Models of \citet{deAvillez_2002} predict that mixing slows down exponentially with separation, which is a possible explanation as to why galaxies are typically found to be less well-mixed on kpc scales than on sub-kpc scales.

\subsection{Metallicity dependence of $\alpha^{\prime}(^{12}{\rm CO})$}

\begin{figure}
    \centering
    \includegraphics[width = 0.5\textwidth, trim={3cm 3cm 3cm 3cm}]{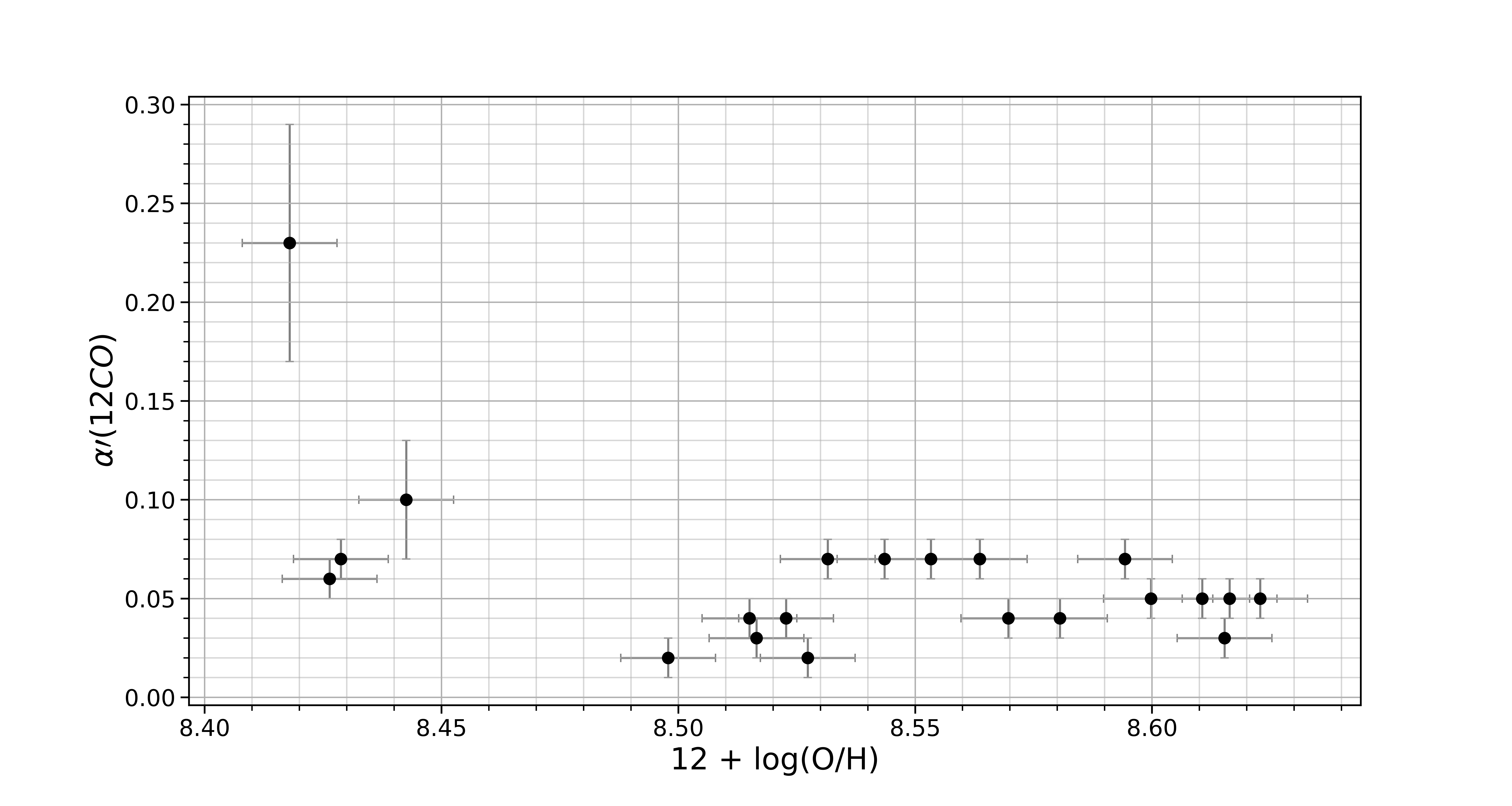}
    \caption{The variation of the CO-to-dust-mass conversion factor, $\alpha^{\prime}(^{12}{\rm CO})$, with metallicity (O) for H~{\sc ii} regions with SMA detections reported by \protect\cite{Viaene_2021}. Some GMCs have multiple associated H~{\sc ii} regions. Uncertainties in the S-calibration-calculated metallicities are the same as for Figure \protect\ref{fig:Ograds}. $\alpha^{\prime}(^{12}{\rm CO})$ values and uncertainties are taken from \protect\citet{Viaene_2021}.}
    \label{fig:alphaco}
\end{figure}

Many of our optical spectra supplement SMA observations for which we have dust continuum emission measurements for individual GMCs (see \citealp{Viaene_2021}). This gives us the opportunity to directly study any apparent metallicity dependence of $\alpha^{\prime}(^{12}{\rm CO})$; the ratio of dust mass and CO luminosity. Both the dust mass and the CO luminosity are likely to vary with metallicity, which could either exacerbate or mitigate ensuing changes of the ratio. What we compare here are \citet{Viaene_2021} $\alpha^{\prime}(^{12}{\rm CO})$ measurements for individual clouds and the metallicities for associated H~{\sc ii} regions obtained in this study. We plot the resulting relationship between H~{\sc ii} region oxygen abundances (from the S-calibration of \citetalias{Pilyugin_2016}) and GMC $\alpha^{\prime}(^{12}{\rm CO})$ (as defined in \citealp{Viaene_2021}) in Figure \ref{fig:alphaco}. Despite the statistically significant difference of the metallicity values, we essentially find a constant $\alpha^{\prime}(^{12}{\rm CO})$ of $\sim (0.0605\pm0.009) M_{\odot}\rm (K~km~ s^{-1}pc^2)^{-1}$ across the metallicity range probed, apart from the two outliers with large uncertainties. Therefore there is no evidence for a strong metallicity dependence of $\alpha^{\prime}(^{12}{\rm CO})$ in this work. This finding may be explained by the prediction from \cite{Bolatto_2013} that a significant change in $\alpha{\rm (CO)}$ is generally only expected for low metallicities i.e. 12 + log(O/H) < 8.4 dex. We see a hint of this trend in Figure \ref{fig:alphaco}, due to the presence of an outlier metallicity at $\sim 8.42$ which has higher $\alpha^{\prime}(^{12}{\rm CO})$ than other sources with more than $1\sigma$ difference. However, this alone is not enough to make a conclusion on a trend of $\alpha^{\prime}(^{12}{\rm CO})$ with metallicity for our sample, a larger sample size spanning the metallicity range 12 + log(O/H) < 8.5 dex is needed to further this particular study. At higher metallicities, this could mean that any impact of metallicity changes in H~{\sc ii} region oxygen abundances and GMC $\alpha^{\prime}(^{12}{\rm CO})$ values largely cancels out. However, we note again that $\alpha^{\prime}(^{12}{\rm CO})$ does not take into account the gas-to-dust ratio, which is also predicted to depend on metallicity.

\section{Summary \& Conclusions} \label{sec:sum}
From our optical spectroscopic survey of M31 we have identified and calculated O and N abundance values for 294 H~{\sc ii} regions from their emission line ratios. From our analysis, we conclude the following:

\begin{enumerate}
    \item We find an Oxygen abundance gradient of $-0.0113 \pm 0.0016 \text{ dex kpc}^{-1}$ by taking the mean of the gradients calculated using the \citetalias{Pilyugin_2016} R-calibration and S-calibration. We explored the use of the R23 diagnostic of \citetalias{Zaritsky_1994}, finding that the value of the radial metallicity gradient depends on the chosen calibration (as has been shown in previous studies e.g. \citetalias{Sanders_2012}), but these generally agree within $2 \sigma$. Weak auroral lines are necessary to calculate a more precise, direct value, but these are only detected in four of our sources with a S/N $>$ 3. 
    \item M31 has a relatively shallow Oxygen abundance gradient compared to other spiral galaxies e.g., those from \citet{Kreckel_2019}. In comparison to the Milky Way it is shallower by a factor of $\sim$ 4 - 6 depending on the choice of methodology.
    \item We find a Nitrogen abundance gradient of $-0.0334 \pm 0.0038 \text{ dex kpc}^{-1}$ from the \citetalias{Pilyugin_2016} R-calibration, in agreement within uncertainty with that from \citetalias{Sanders_2012} calculated for M31 using the diagnostic from \cite{Pilyugin_2010}. 
    \item The N abundance gradient of M31 is $\sim$ 1.7 - 3.4 (depending on the chosen calibration) times steeper than the O abundance gradient, this indicates that the N/O ratio is higher towards the centre. This is consistent with cosmological simulations \citep{Vincenzo_2018} which show a negative N/O radial gradient.
    \item Around our S-cal O radial gradient we measure a standard deviation in scatter of $\sim$ 0.06 dex, exceeding the mean inherent measurement uncertainty of $\sim$ 0.01 dex. After subtracting the radial gradient from metallicities to get a residual metallicity, $\Delta(\rm O/H)$, we see significant outliers that exceed this scatter up to 5$\sigma$. This confirms conclusions from previous works (e.g., \citetalias{Sanders_2012}, \citealp{Kreckel_2020}) that a radial gradient is not sufficient to describe metallicity variation between H~{\sc ii} regions in spiral galaxies.
    \item The presence of reduced-metallicity (compared to the radial gradient) H~{\sc ii} regions throughout the disk is possible evidence that external gas of lower relative metallicity has been brought into the galaxy, possibly from outside. This may have occurred during a collision between M31 and M32 occurring 200 - 800 Myrs ago. We also see the additional presence of enhanced-metallicity (compared to the radial gradient) H~{\sc ii} regions throughout the disk, which previous work suggests is due to ISM enrichment by recent star-formation and supernovae \citep{Ho_2017}. 
    \item From the two-point correlation function, we see that metallicities are well-correlated (compared to the random sample) at sub-kpc scales. There is evidence of an exponential decrease in the speed of mixing (i.e. \citealp{deAvillez_2002}) and so it is expected that kpc-scale inhomogenities will be present for longer periods of time. Additionally, inefficient mixing on larger scales could suggest that not all material brought in by the collision, if this occured >$\sim$ 800 Myrs ago, is mixed yet and causes reduced abundance H~{\sc ii} regions in our sample.
    \item We study the variation in $\alpha^{\prime}(^{12}{\rm CO})$ from an SMA survey of the associated GMCs \citep{Viaene_2021} with O abundance. Despite a relatively large range in metallicity, we find no evidence for a significant dependence of $\alpha^{\prime}(^{12}{\rm CO})$ on O abundance at 12 + log(O/H) > 8.42 for H~{\sc ii} regions associated with GMCs observed by the SMA. The data point in Figure \ref{fig:alphaco} with metallicity $\sim 8.42$ has a respectively large error bar in comparison to other sources, suggesting that this high $\alpha^{\prime}(^{12}{\rm CO})$ value is not statistically significant.
    \item In future work we are aiming to obtain $\alpha^{\prime}(^{12}{\rm CO})$ measurements for more GMCs in M31 for which we also have H~{\sc ii} region abundances, so that the correlation between the two factors can be investigated further. We will also investigate the chemical abundance throughout the M31 disk further through analysis of log(N/O).
\end{enumerate}
\section*{Acknowledgements}
The authors thank the anonymous referee for helpful comments.
Observations reported here were obtained at the MMT Observatory, a joint facility of the Smithsonian Institution and the University of Arizona.
C.K. acknowledges funding from the UK Science and Technology Facility Council (STFC) through grants ST/Y001443/1.

\section*{Data Availability}
H~{\sc ii} region spectra can be shared upon reasonable request.


\bibliographystyle{mnras}
\bibliography{example} 



\appendix

\section{Emission Line Profiles}
\label{sec: LineProfs}
\begin{figure*}
    \centering
    \subfigure{
        \includegraphics[width=0.9\textwidth, trim={4cm 3cm 4cm 2cm}]{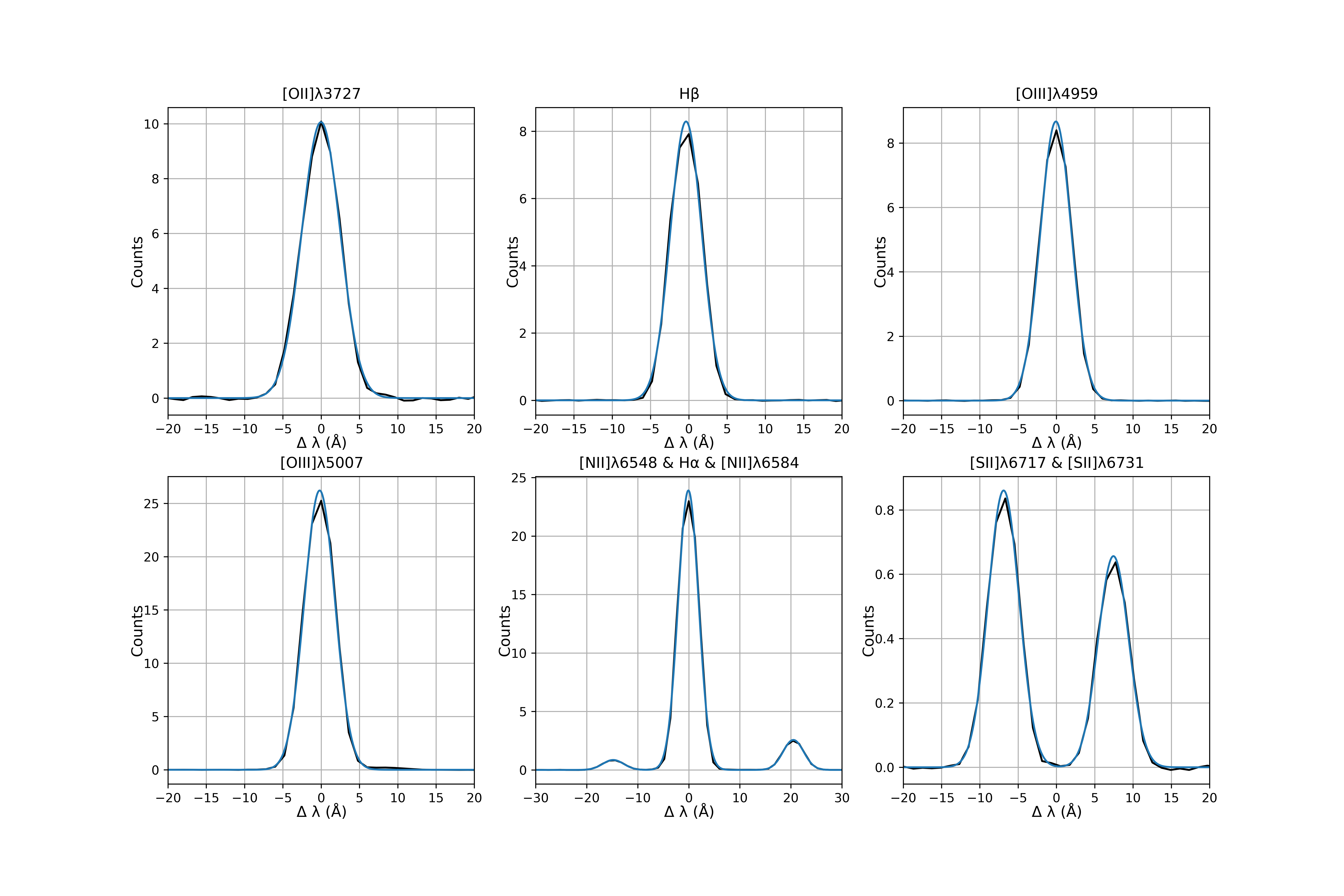}}   
        
        \vspace{1em}
        \vspace{1em}
        
    \subfigure{
        \includegraphics[width=0.9\textwidth, trim={4cm 3cm 4cm 2cm}]{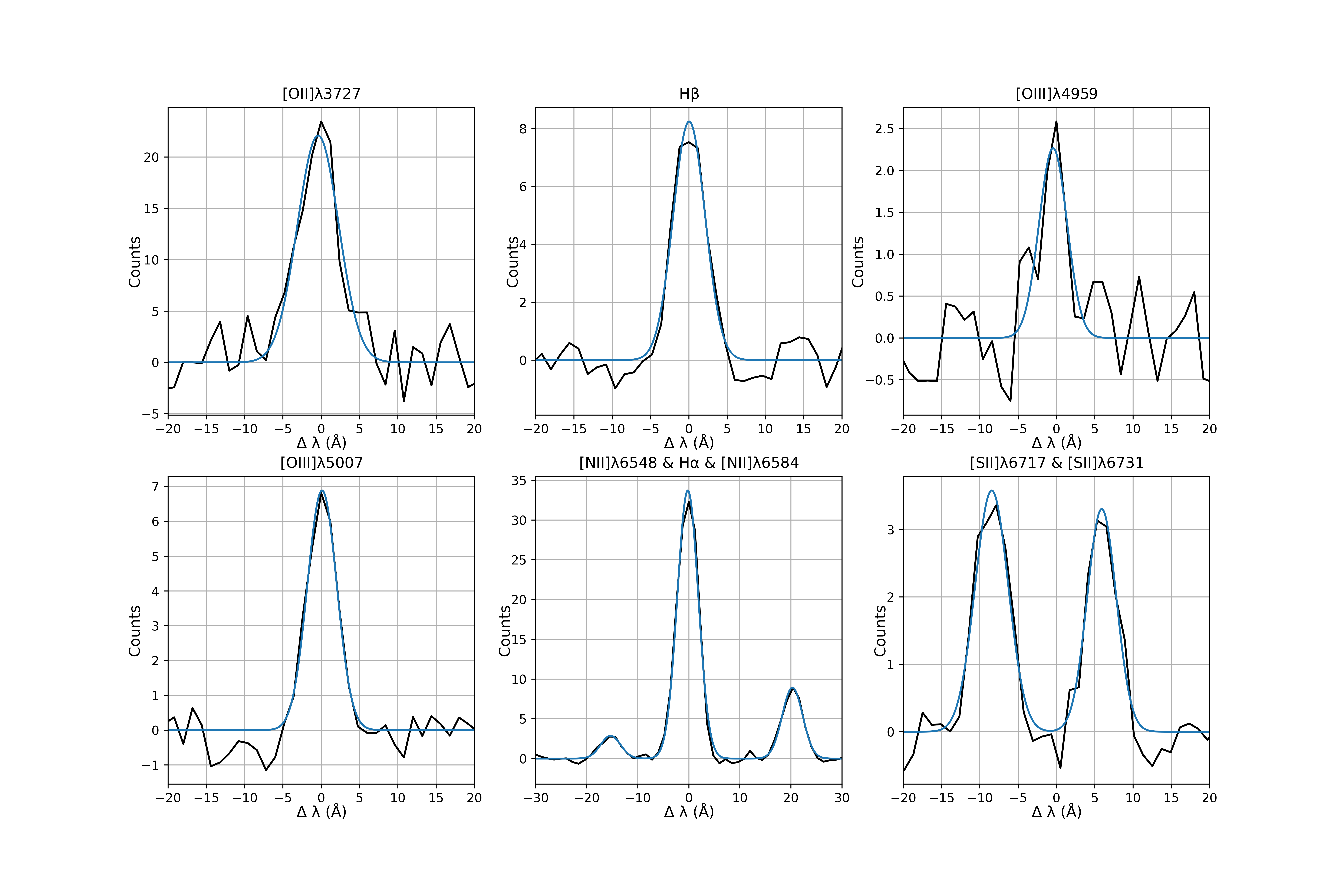}}
    
    \caption{Emission line profiles for the strong lines measured in two example spectra. TOP: Spectrum in the top 2\% signal-to-noise (row 7 from the data tables). BOTTOM: Spectrum in the bottom 30\% signal-to-noise (row 6 from the data tables). On the x-axis, 0 represents the rest-wavelength of the lines. In the case the line profile of $\rm H\alpha$ and the [NII] doublet, we centre the wavelength-axis on the midpoint of $\rm H\alpha$. For the [SII] doublet, we centre the wavelength-axis on the midpoint between the rest wavelengths of both lines.}
    \label{fig:LineProfs}
\end{figure*}

As an example, we show emission line profiles for the strong lines measured (see Section \ref{sec: ELF}) to calculate Oxygen and Nitrogen abundances (see Section \ref{sec:met}) for two spectra from M31 H~{\sc ii} regions in our sample in Figure \ref{fig:LineProfs}. One spectrum from the top 2\% highest signal-to-noise spectra in our sample, corresponding to row 7 in Tables \ref{tab:tab_fluxes} \& \ref{tab:tab_met}, and one spectrum from the bottom 30\% lowest signal-to-noise spectra in our sample, corresponding to row 6 in Tables \ref{tab:tab_fluxes} \& \ref{tab:tab_met}. The respective Gaussian fits to each line profile are additionally shown.

\twocolumn


\bsp	
\label{lastpage}
\end{document}